\documentstyle[12pt,aaspp4,flushrt]{article}

\begin{document}

\title{The Las Campanas Redshift Survey}

\author{Stephen A. Shectman and Stephen D. Landy}
\affil{Carnegie Observatories, 813 Santa Barbara St., 
       Pasadena, CA 91101 USA \\
       shec@ociw.edu, landy@ociw.edu}

\author{Augustus Oemler}
\affil{Dept. of Astronomy, Yale University, 
       New Haven, CT 06520-8101 USA \\
       oemler@astro.yale.edu}
\author{Douglas L. Tucker}
\affil{Astrophysikalisches Institut Potsdam, An der Sternwarte 16,
       D-14482 Potsdam, Germany \\
       dtucker@aip.de}
\author{Huan Lin \altaffilmark{1} and Robert P. Kirshner}
\affil{Harvard-Smithsonian Center for Astrophysics, 60 Garden St.,
       Cambridge, MA 02138 USA \\
       hlin@cfa.harvard.edu, kirshner@cfa.harvard.edu}
\and
\author{Paul L. Schechter}

\affil{Dept. of Physics, Massachusetts Institute of Technology,
       Cambridge, MA 02139 USA \\
       schech@achernar.mit.edu}
\altaffiltext{1}{Present Affiliation: Dept. of Astronomy, University of
    Toronto, 60 St. George St., Toronto, ON M5S 3H8, Canada, 
    lin@astro.utoronto.ca}

\begin{abstract}
The Las Campanas Redshift Survey (LCRS) consists of 26418 redshifts of galaxies
selected from a CCD-based catalog obtained in the $R$ band.  The survey
covers over 700 square degrees in 6 strips, each 1.5$\arcdeg$ x 80$\arcdeg$,
three each in the North and South galactic caps.  The median redshift in the 
survey is about 30 000 km~s$^{-1}$.  Essential features of the
galaxy selection and redshift measurement methods are described and 
tabulated here.  These details are important for subsequent analysis
of the LCRS data.  Two dimensional representations of the redshift 
distributions reveal many repetitions of voids, on the scale of about
5000 km~s$^{-1}$, sharply bounded by large
walls of galaxies as seen in nearby surveys.  Statistical
investigations of the mean galaxy properties and of clustering on the large
scale are reported elsewhere.  These include studies of the luminosity
function, power spectrum in two and three dimensions, correlation function, 
pairwise velocity distribution, identification of large scale
structures, and a group catalog.  
The LCRS redshift catalog will be made
available to interested investigators at an internet web site and in archival
form as an Astrophysical Journal CD-ROM.

\end{abstract}

\keywords{cosmology: observations --- galaxies: clustering --- 
          galaxies: distances and redshifts --- 
          galaxies: luminosity function, mass function --- 
          large scale structure of universe}


\section{Introduction}

Redshift surveys reveal surprising structures in the
large-scale distribution of galaxies, which provide clues to 
physical properties of the Universe (see \cite{gio91} and
\cite{strwil95} for reviews.)  
Early investigations (such as \cite{KOSSI}, or \cite{huchra83})
suggested that the nearby Universe might be strongly inhomogeneous 
with nearly empty voids and thin, high contrast regions of galaxy overdensity.
Convincing demonstrations that this pattern is a
general feature of the Universe have come from extensive surveys which 
cover large angles on the sky, (\cite{deL86}, \cite{gel89}:summarizing
the CfA surveys, and \cite{dacosta94a}: the SSRS2 survey.) 
Galaxies appear to lie on networks of filaments or sheets extending over 
$100 \ h^{-1}$~Mpc that encompass sharply bounded voids
(Hubble constant $H_0 = 100 \ h$~km~s$^{-1}$~Mpc$^{-1}$).
Despite the power of the CfA2 (\cite{huchra95}) and SSRS2 surveys in 
conveying an impression of the galaxy distribution on large scales,
the largest features they reveal are very near the upper limit set by the
depth of the survey at about 12000~km~s$^{-1}$.  

Are these the largest features in the
Universe?   There are hints of possible structure on larger scales from 
apparent periodicities in redshifts (\cite{broadhurst}), from the
flows suggested by 
Lauer \& Postman (1994), or from possible variations in the luminosity
density (\cite{dacosta94a}).
Only a deep extensive survey can provide evidence on the reality of
these suggestions.  Similarly, the statistical measures of galaxy clustering
derived from redshift catalogs (\cite{Efstathiou}, \cite{park})\ which
can be used to constrain models for the formation of structure in the Universe,
have only modest precision at the large scales which are most interesting.
To explore structure in the local Universe on a scale of
30000~km~s$^{-1}$ and to improve our measures of galaxy statistics, we
have been working since 1987 on the
Las Campanas Redshift Survey (LCRS).  We have compiled a CCD-based
galaxy catalog, selected the galaxies, developed the fiber-optic
equipment for mass producing
galaxy spectra, and observed from 1988 to 1994.  We have measured more than 
26000 galaxy redshifts averaging $z = 0.1$ over an area of 700 square degrees
arranged in six long thin strips, three in the North Galactic Cap, and
three in the South.  The Las Campanas Redshift Survey provides a 
reconaissance of present-day structure on the largest scales mapped to date. 

The Las Campanas Redshift Survey is a direct descendent of
earlier surveys aimed at measuring the average properties of galaxies:
the luminosity function and the space density of galaxies
(\cite{KOSI}; \cite{KOSII}; \cite{KOSSII})  Annoying variations in the average
properties, such as the luminosity density, signalled the 
presence of strong inhomogeneities in the
galaxy distribution. Most startling was a large void of diameter 
$60 \ h^{-1}$~Mpc in the galaxy 
distribution beyond the constellation Bo\"{o}tes
(\cite{KOSSI}; \cite{KOSSIII}; \cite{KOSSIV}). Because similar
structures are now frequently
seen in wider-angle surveys and because the structures are nearly as large as 
the survey dimensions, a much larger and deeper 
survey seems needed in order to encompass a fair sample of the universe,
as well as to search for structure on larger scales.
The LCRS began with a goal of 
obtaining a galaxy sample of sufficient size, sky coverage, and depth
to permit a reliable characterization of the average properties of
galaxies and of their distribution. 

While this work was in progress, some important large wide-angle
redshift surveys have been carried out.
Recent large-scale structure analyses include (see \cite{strwil95} for a more 
complete list): (1) nearly all-sky samples of
objects selected from the IRAS Point Source Catalog, 
specifically the IRAS 1.2 Jy survey (5321 galaxies with 60~$\mu$m flux
$f_{60} > 1.2$~Jy; \cite{fisher95}) and the 1-in-6 sparse-sampled
IRAS QDOT survey (2184 galaxies with
$f_{60} > 0.6$~Jy; \cite{law95}); (2) the Stromlo-APM
survey, selected from the optical APM galaxy catalog, covering $4300$ square
degrees of the Southern sky to a median survey depth 
$cz = 15000$~km~s$^{-1}$, but
1 in 20 sparse sampled (1787 galaxies, $b_J \leq 17.15$; 
\cite{1:lov92}; \cite{1:lov95}; \cite{lov96});
and (3) the combined CfA2 and Southern Sky (SSRS2) redshift surveys, 
selected from the Zwicky catalog and from plate scans, respectively,
covering one-third of the sky over the North and South galactic caps out to
a mean depth of about $7500$~km~s$^{-1}$ 
(14383 galaxies with $m_{B(0)} \leq 15.5$; 
\cite{huchra95}, \cite{dacosta94a}, 1994b). The CfA2+SSRS2 sample
represents the state of the art for a wide-angle survey, with a large
sample size and dense sky coverage.  Sparse sampling provides an efficient 
way to measure
a particular statistic, while dense sampling reveals structures 
that require the coherent arrangement of
many galaxies and which are not easily characterized
by low-order statistics.  Our survey has some of the desirable properties of each:
the samples are strips on the sky 1.5$\arcdeg$ wide and  80$\arcdeg$ long which are
separated by 3$\arcdeg$. Within each strip, the galaxies are sampled
randomly from a magnitude-limited catalog, but the sampling is quite dense,
averaging about 70\% of the magnitude-limited list.  The distinctive properties of our
survey are its depth (out to 60000 km~s$^{-1}$) and the number of redshifts (26000).     

The Las Campanas Redshift Survey provides improvements in sample size, volume, and
depth in a limited region of the sky. Unlike most previously completed surveys, 
where spectra were
taken of individual galaxies, one at
a time, the LCRS employs a fiber-optic spectrograph system on
the Las Campanas 2.5 meter DuPont telescope to observe over 100 objects at once. 
The greater depths explored
by the LCRS mean that we observe galaxies which have sufficient surface 
density on the sky to make efficient
use of a multi-fiber system.  Some details of the galaxy selection and observing 
procedure which add to the complexity of the analysis were shaped by the 
properties of the fiber system, but
the overall gain for exploring large-scale structure is very large. 
In this sense the LCRS represents part of a fundamental technological 
change in the way that large-scale redshift surveys are carried out.

Other surveys in progress, such as the Century Survey (\cite{century}) and the
ESO-based survey (\cite{eso}), have a comparable depth but cover smaller areas. 
Each has selection
criteria that differ from ours in interesting ways which should make the 
comparison of our results with theirs a fruitful enterprise.
The next generation of surveys, such as the 2 Degree Field project 
(\cite{ellis93}) and 
the the Sloan Digital Sky Survey (\cite{gunwein95}), will use
the multiplex advantage of fiber-fed spectrographs to obtain redshifts
of up to a million galaxies, to about the same depth ($z \approx 0.1$) 
explored by the LCRS.  It is our hope to illuminate some of the most 
interesting subjects for more thorough investigation by these ambitious projects.

In addition to providing a map of the galaxy distribution on the largest scales, 
the LCRS can help to constrain the 
physical history of the Universe and the properties of its constituents.  
Current theories of structure formation start
with a spectrum of fluctuations produced in the early universe, 
follow the growth of 
structure due to gravitation, and strive to
match both the subtle structure observed at recombination and the 
high-contrast distribution of the galaxies today mapped by redshift data.  
Most models 
presume that the Universe is dominated by non-baryonic dark matter (\cite{blum84})
whose properties are to be inferred from the match to the data.
The luminous galaxies trace the 
underlying mass, but may be biased:  galaxies may form preferentially
in peaks of the matter distribution (\cite{bard86}, \cite{white87}).
Such models, with the aid of numerical N-body simulations, predict
the clustering properties of galaxies from small nonlinear scales of
less than $10 \ h^{-1}$~Mpc, all the way up past the largest scales sampled 
by existing redshift surveys.
The scales sampled with the LCRS, up to about $300 \ h^{-1}$~Mpc, correspond 
to the horizon size at interesting stages in the 
early universe and are closer to the scales probed by microwave backgound 
observations than can be measured by smaller surveys. By observing galaxy fluctuations
on these scales with LCRS, we can learn about primordial density fluctuations
and the processes by which they grow.  Combining evidence from large-scale 
structures 
and from microwave
background observations holds the potential to answer fundamental
questions about the dark matter that makes up 
most of the universe (\cite{scott95}).

This paper gives details about
the construction of the LCRS which are essential to any analysis of the
survey data.  Some preliminary results and descriptions of the survey 
have already been published (\cite {she92}, \cite {oem93}, \cite {mapscale}, 
\cite {moriond}, \cite {she95}, \cite {tuc95}.)
Tucker's (1994) Ph.D. thesis at Yale discusses the construction of the 
survey's photometric catalog and statistical analyses of the early 
50-fiber sample.  Tucker describes
the galaxy correlation function, the properties of galaxy
groups, and galaxy clustering as a function of galaxy color.  A more
complete analysis, using the entire LCRS data set to investigate these
topics, will be carried out by Tucker (Tucker et al.\ 1996c,a,b, 
respectively). 
Huan Lin's Ph.D. thesis at Harvard (\cite {thesis}) is the basis for
the present paper and for several forthcoming papers on the LCRS.
Lin et al. (1996a) details the 
derivation of the luminosity function and space density of LCRS galaxies. 
Lin et al. (1996b) deals with the power spectrum,
a basic statistic which characterizes the clustering properties of 
galaxies.  Lin et al. (1996c) examines redshift-space 
distortions in the 
clustering of LCRS galaxies, derives the velocity dispersion of galaxy pairs,
and estimates the value of the cosmological density parameter $\Omega$.  
The analysis of Landy et al. (1996) uses the
tools of 2D Fourier analysis to search for coherent structures on very large
scales in the LCRS data.  Doroshkevich et al. (1996) examines the
typical scales and types of 
large scale structure in the LCRS sample.
Our intention is to make results from this 
survey and the survey data itself available to
interested investigators.  We have established a home page for the LCRS
at ``http://manaslu.astro.utoronto.ca/\~{ }lin/lcrs.html'' which
will provide rapid access to the data.  An archival table of the
redshift data will accompany this paper as an Astrophysical Journal CD-ROM.

\section{Survey Construction}
 
The Las Campanas Redshift Survey has been carried
out at the Carnegie Institution's Las Campanas Observatory in Chile, 
with redshifts harvested in a series of 13 spectroscopic observing 
runs from November 1988 
through October 1994.
Over 26000 galaxy spectra have been gathered, with an average redshift
$z = 0.1$, over 700 square degrees of the sky.
This survey has exploited the Las Campanas 2.5-m DuPont telescope's
$2.1\arcdeg$-diameter field and the multiplexing capability of the 
associated multi-object fiber-optic spectrograph system constructed by 
Shectman, which permits simultaneous
observations of 112 galaxy spectra.

The survey data may be divided into two parts based on the development
of the photometric detectors and the fiber system during the six year span 
of the LCRS. The first 20\% of the 
survey galaxies were selected based on data from an 800 x 800
TI CCD and the
redshifts were measured using a 50-object fiber system. Beginning 
October 1991, an improved 112-object system was used to gather the
final 80\% of the data, which were also selected with better imaging
systems using first a LORAL, then a large Tektronix CCD. See 
Table~\ref{tabnum} for a summary. 
Some significant technical details of the survey construction, and additional 
specifics of the instruments and methods 
may be found in
Shectman et al.\ (1992, 1995) and Tucker (1994), but essential features
which affect the use of the sample for further analysis are set out here.

\subsection{Photometry, Astrometry, and Object Selection}

There was no suitable galaxy catalog available to us in 1986 when preparations 
for the Las Campanas survey began in earnest.  We constructed our own catalog
through a
modest digital sky survey at the 1m Swope Telescope at Las Campanas.  
Photometry for the survey comes from  CCD drift scans, taken
through a Thuan-Gunn $r$ filter (\cite{thu76}) with the telescope's
sidereal drive turned off.  The accumulated charge is 
shifted across the chip at the sidereal rate and read out to produce
a continuous strip of photometric data.  Three CCD's have been used in this
way since the
beginning of the survey.  The first was a thinned 800 x 800 TI device, 
which was used
with reimaging optics to obtain 1$\farcs$06 pixels.  This was replaced in 1990 with a 
thick 2K x 2K LORAL chip, used without reimaging optics and binned 2 x 2 to provide
0$\farcs$865 pixels.  The third CCD, introduced in 1992, is a 
thick 2K x 2K Tektronix CCD 
used unbinned without
reimaging to provide 0$\farcs$692 pixels.  Star-galaxy separation is appreciably better
with the latter two chips because of the increased scale and the absence of
distortion from intermediate optics.

The effective exposure time depends on the
size of the chip and the cosine of the declination, but was typically 
near 1 minute.  The limiting
magnitude for the scans is similar to the limiting magnitude of the 
ESO red plates of the
Southern Sky survey and extended well below the adopted limits near 
$m = 17.75$ used
for galaxy selection.  Overlapping scans, offset in declination by
slightly less than a chip width,
were obtained to build up a strip $1.5\arcdeg$
wide.  The typical length of each scan was $3\arcdeg$ 
or $4.5\arcdeg$ long: short enough
to complete each $1.5\arcdeg$ x $3\arcdeg$ ``brick'' in about 2 hours, 
but long enough to 
minimize the time lost in setting the telescope and initiating the scan.  
Each brick covers two or three $1.5\arcdeg \times 1.5\arcdeg$ 
spectroscopic fields for the fiber system described below.
A set of bricks
was accumulated over several observing seasons to pave 700 square degrees, 
arranged in six long paths: in the North galactic hemisphere 
at $\delta = -3\arcdeg, -6\arcdeg, -12\arcdeg $
and in the South galactic hemisphere 
at $\delta = -39\arcdeg, -42\arcdeg$, and $-45\arcdeg $.

\begin{figure}
\plotfiddle{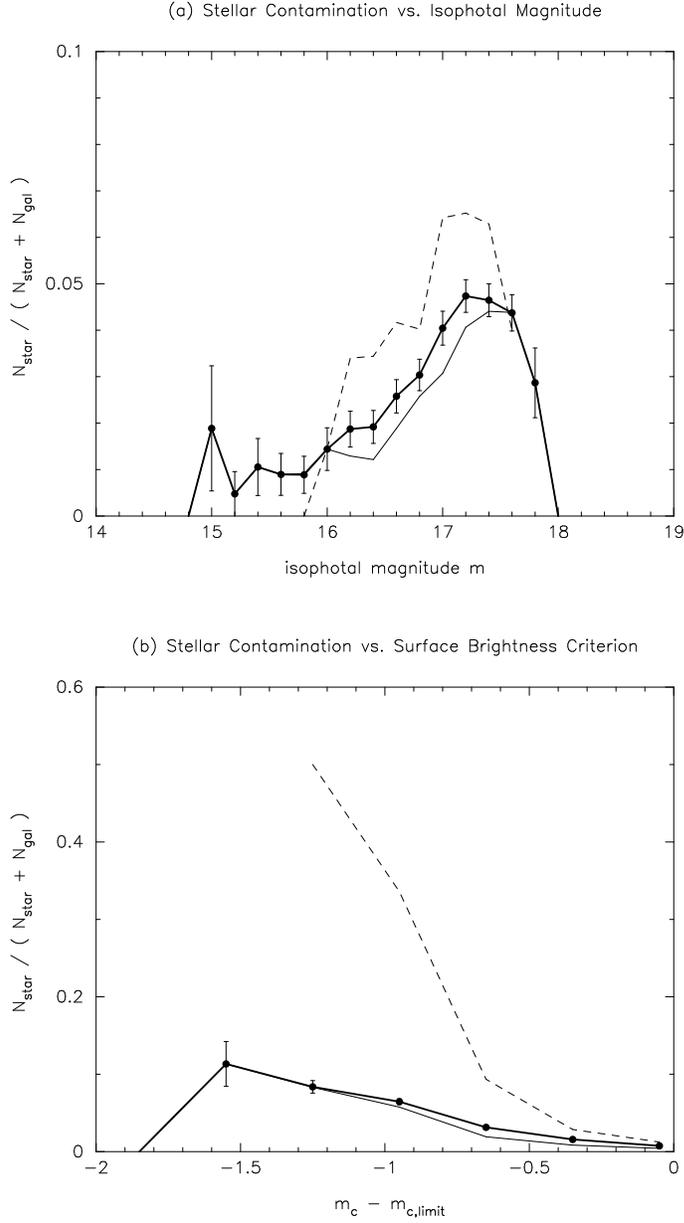}{17cm}{0}{80}{80}{-250}{-100}
\caption{The ratio $N_{star} / ( N_{star} + N_{gal} )$ vs. (a) isophotal
    magnitude and (b) vs. distance $m_c - m_{c,limit}$ from the
    central surface brightness ``cut line,'' 
    where $m_{c,limit} \equiv m_{cen} - 0.5(m_2 - m)$.
    Here, $m_{cen}$ 
    and $m_2$ are the proper values for an object within a given field and are
    taken from Table~2.  In both plots, the filled circles
    connected by the 
    thick solid line denote the ratio for the full LCRS sample; the dashed 
    line, the 50-fiber sample; and the thin solid line, the 112-fiber sample.
    Note that, in (b), the high value of the ratio for the 50-fiber sample
    at $m_c - m_{c,limit} = -1.25$, is due to just two objects -- one
    star and one galaxy. Poisson errors are shown.}
\label{figstar}
\end{figure}

These drift scans were analyzed by an automated photometry system
similar to that described in Kirshner et al. (1983). 
Bias was determined from an overscan region and subtracted from
the data.  The data were then sky-subtracted and flatfielded using 
the mode of the pixel values in each column and in each scan-line,
respectively.
Objects were found by a search routine which identified
contiguous groups of pixels whose brightness was greater than 1.15
times that of the sky. Two magnitudes were derived for each object: an
isophotal magnitude, $m$, corresponding to the sum of the background
corrected flux in all pixels within the object, and a central
magnitude, $m_c$, which measures the flux within a 2-pixel radius of
the object center. The area measured by the central magnitude is 
close to that covered by a spectroscopic fiber and it is used to
identify galaxies which have low central surface brightness. Images were classified using 
several structural parameters. All objects not classified as stars
were inspected by eye to eliminate spurious galaxy classifications. 
Our star-galaxy separation criteria turned out to be conservative so
that there is a small amount of stellar contamination in the
spectroscopic sample (Table~\ref{tabnum} and \S~\ref{spec}). 
Figure~\ref{figstar} illustrates the stellar contamination rate as a
function of isophotal magnitude and surface brightness.

\begin{figure}
\plotfiddle{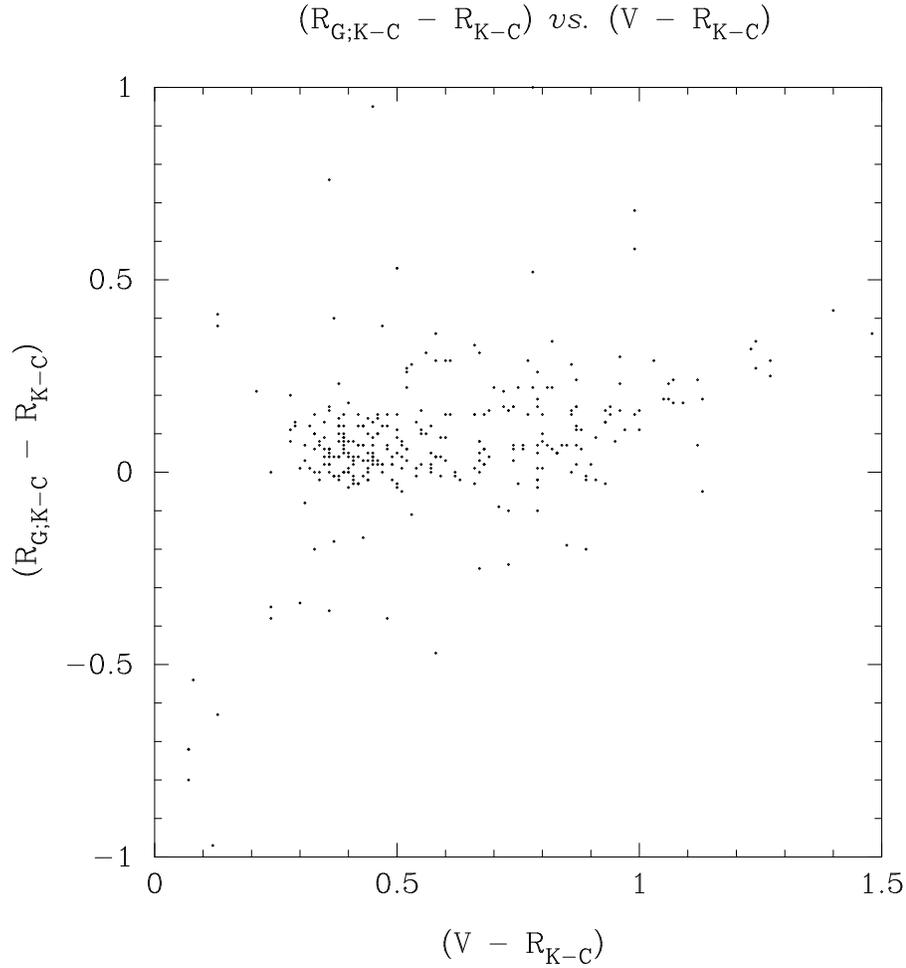}{17cm}{0}{80}{80}{-275}{-100}
\caption{The difference between the LCRS hybrid Kron-Cousins $R$, $R_{G;K-C}$, 
    and true Kron-Cousins $R$, $R_{K-C}$, vs. $(V - R_{K-C})$ color.  Note that
    the systematic difference between the two $R$ bands is less than 0.1 mag.
    (True Kron-Cousins $R$ magnitudes for LCRS objects were obtained from
    calibration frames taken at the CTIO 0.9 meter telescope.)}
\label{fig2-4}
\end{figure}

Photometric calibration of the data proceeded in several
steps. Observations were obtained only on nights which appeared to be
photometric, and photometric standard fields were observed several
times per night. These standards were used to obtain a zeropoint for
each night's scans. 
Though our photometry was obtained with a Gunn $r$ filter, the 
calibration was done relative to standards (\cite{gra81};
\cite{gra82}) with magnitudes in the Kron-Cousins $R$ band, 
resulting in a hybrid red band which we call
$R_{G;K-C}$.  The zeropoint difference between $R_{G;K-C}$ and true
Kron-Cousins $R$ is small ($< 0.1$ mag), which can be seen in Figure~\ref{fig2-4}
(see also \cite{tuc94}).

This photometry calibrated the catalog from
which objects were selected for spectroscopic observations.  Even after the 
observations were obtained, we continued to refine
the photometric calibration, so the resulting photometric limits 
for each brick vary slightly from the nominal values. Near the
end of this survey, after scans were obtained of all
bricks in each declination strip, additional calibration procedures
were applied to eliminate residual photometric offsets. These offsets
can be due to two effects. First, fluctuations in the derived
zeropoints from night to night, and within some nights, reveal that
not all of the nights on which data were obtained were perfectly
photometric. Secondly, isophotal magnitudes of galaxies are quite
sensitive to the limiting isophote. Because our outer isophote was 
defined as a 
multiple of the sky brightness, variations in sky brightness could cause
variations in $m_{iso}-m_{total}$, as could variations in the seeing. 

Because most of the bricks in a strip overlap their neighbors to the
east and west,
we could compare the photometric zeropoints of all the bricks
within one, or at most a few, groups of contiguous bricks within each strip. This
comparison, using the stars in the
overlapping regions, revealed residual photometric offsets with a
standard deviation of 0.065 mag. After removing these offsets, 
the zeropoint of each group of bricks was redetermined using
CCD observations of 26 regions we obtained with 
Yale Service Observing Time on 
the CTIO\footnote{Cerro Tololo Inter-American Observatory, National Optical 
    Astronomy Observatories, which are supported by the Association of
    Universities for Research in Astronomy, Inc., under cooperative
    agreement with the National Science Foundation.} 
0.9 meter telescope, supplemented with objects from the HST
photometric catalog. This comparison revealed no offsets in the mean of
any group greater than $\sim 0.015$ mag. 

To remove variations in $m_{iso}-m_{total}$, magnitudes were
calculated for all objects through a 12 arcsec diameter synthetic aperture. This
magnitude, $m_{12}$, is insensitive to seeing or sky brightness
variations. Therefore, $m_{12} - m_{total}$, while a function of
magnitude, should be constant from field to field.  Intercomparison of
$m_{iso}-m_{12}$ with $m_{iso}$ from field to field allows us
to remove variations in $m_{iso}-m_{total}$ for galaxies. The
field-to-field variation in this quantity had a standard deviation of
0.07 mag.  

\begin{figure}
\plotfiddle{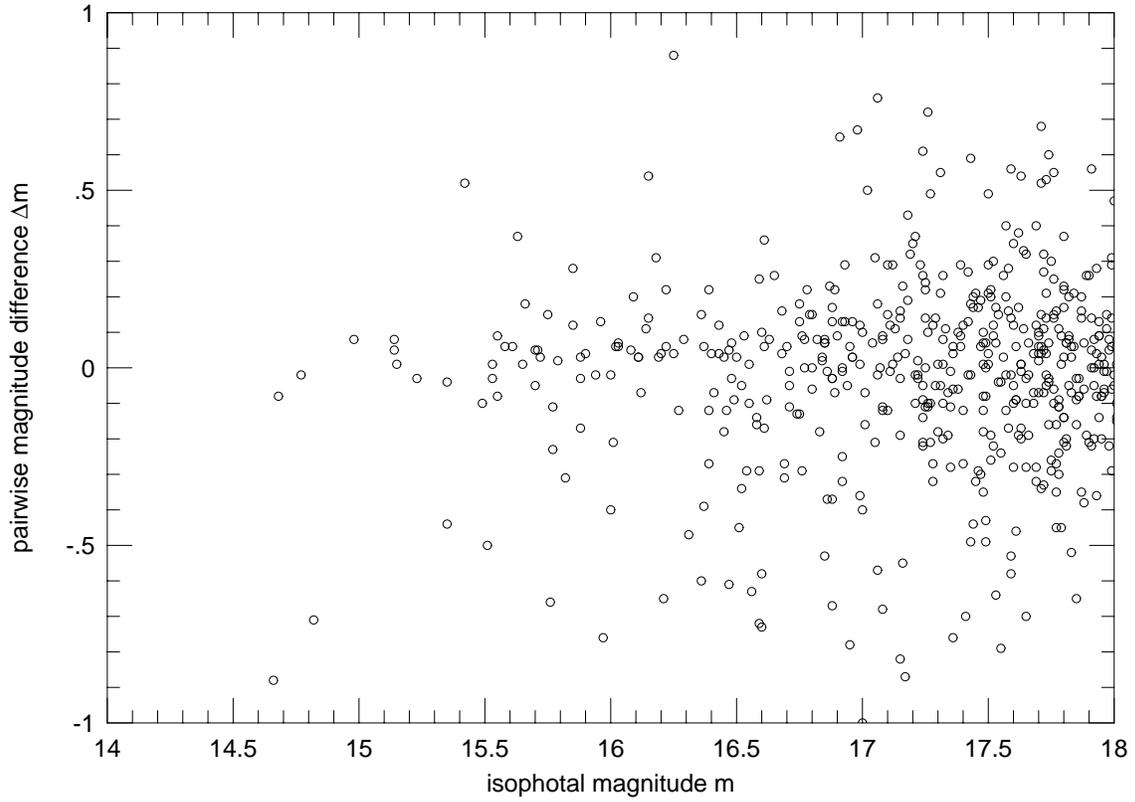}{13cm}{0}{62}{62}{-250}{50}
\caption{The final distribution of pairwise galaxy magnitude differences
   for galaxies which were measured twice on overlapping bricks of
   the LCRS.}
\label{figmag}
\end{figure}

The final distribution of pairwise galaxy magnitude differences
for galaxies which were measured twice on
overlapping bricks is presented in Figure~\ref{figmag}. 
This distribution
has broader wings than expected from pure photon noise, but is
otherwise as expected. From this distribution we calculate the
standard deviation of galaxy magnitudes, $\sigma_m = 0.10$, for $16.0
< r \le 17.0$, and $\sigma_m = 0.17$ for $17.0 < r \le 18.0$. 

The astrometric calibration of driftscan data is straightforward. To
first order, the $x$ and $y$ coordinates within the data array map
linearly into declination and right ascension, with two scale factors:
telescope scale and CCD clock rate. Three additional correction factors
may need to be applied: one for rotation of the CCD relative to the
cardinal directions of celestial coordinates, a second for differential stellar
aberration as the earth rotates during a scan, and, in the case of the
early data taken with a focal reducer, a third for distortions in the optical
system. Stellar aberration was calculated from first
principles. Experience revealed that the other factors were all stable
during a single observing run, and one global calibration of each was
sufficient per run. 

This calibration was done using the coordinate system defined by the HST Guide
Star Catalog (\cite{jen90}, \cite{las90}, \cite{rus90}). 
The Guide Star Catalog has the virtue of containing a
very dense sample, although individual positions are not very accurate (of
order 1 arcsec) and it suffers from systematic errors of up to
several arcsec near the edge of the individual Schmidt telescope
fields from which the positions were derived. When averaged over all
of the bricks observed in a single run, these errors are unimportant
for determining the calibration parameters for the run. They are not,
however, negligible for determining the coordinate zeropoints of each
brick. When adjustment is made for systematic offsets between
overlapping bricks, we find that the pairwise position differences
between the bricks implies a standard deviation in the 1 dimensional
position of an object $\sigma_{\alpha\delta}$ = 0$\farcs$25, for both
stars and galaxies at all magnitudes in our catalog. However, because of the
systematic zeropoint uncertainties, the absolute coordinates should
not be trusted to better than 1 arcsec.

\begin{figure}
\plotfiddle{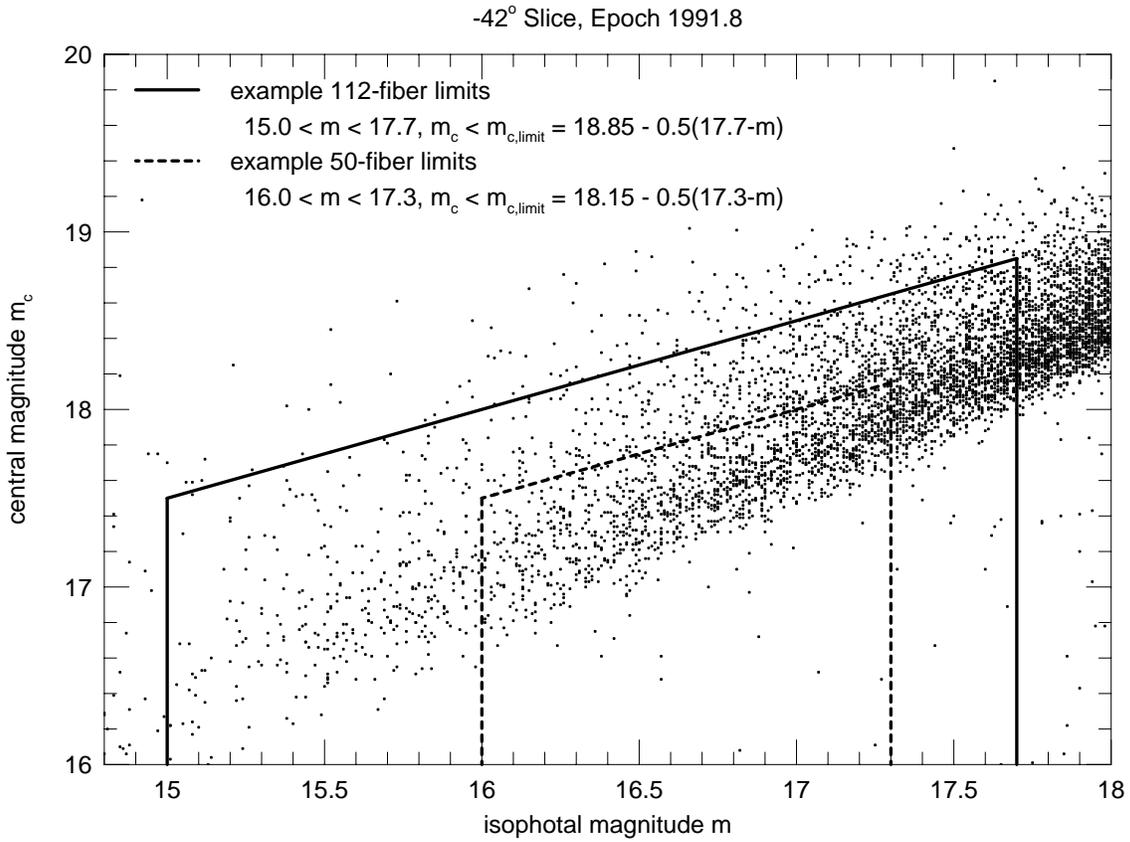}{13cm}{0}{62}{62}{-250}{0}
\caption{Examples of the photometric selection criteria in the
   isophotal magnitude ($m$) - central magnitude ($m_c$) plane
   applied to the 112- and 50-fiber LCRS data.}
\label{figsel}
\end{figure}

Two criteria are applied to select objects, classified as galaxies 
from the photometric catalog, 
for follow-up spectroscopy. First, both faint and bright isophotal magnitude
limits are applied, with the following nominal values:
\begin{equation}
m_1 \leq m < m_2 :
 \left\{
  \begin{array}{ccc}
    m_1 = 16.0, & m_2 = 17.3, & {\rm 50  \ fiber \ data} \\
    m_1 = 15.0, & m_2 = 17.7, & {\rm 112 \ fiber \ data}
  \end{array}
 \right. \ .
\end{equation}
Second, the following nominal central surface brightness ``cut line'' 
is imposed:
\begin{equation}
m_c < m_{cen} - 0.5 (m_2 - m):
 \left\{
  \begin{array}{ccc}
    m_{cen} = 18.15, & {\rm 50  \ fiber \ data} \\
    m_{cen} = 18.85, & {\rm 112 \ fiber \ data}
  \end{array}
 \right. \ .
\end{equation}
Figure~\ref{figsel}
illustrates these selection criteria.
Although they are more complex than a simple
magnitude limit, they have their origin in the technique of our
fiber optic measurements.  Other surveys also have
limits in surface brightness, but they are rarely made explicit:  they
result from failed observations.
The faint isophotal limit $m_2$ is chosen so that there would generally 
be slightly more galaxies than fibers in each spectroscopic field.  
The bright isophotal limit $m_1$ is used so that the total count rate
through the fibers, used to guide telescope pointing during the spectroscopic
exposures, is not dominated by a few bright galaxies.
The cut line limit on $m_{c}$ is used to eliminate low surface 
brightness (LSB) galaxies, 
for which
it is difficult to obtain redshifts during the fixed exposure
time of two hours. The slope of the line is chosen so that the 
fraction of LSB galaxies
eliminated is approximately constant with isophotal magnitude, about 20\%
for the 50-fiber data, but $< 10\%$ for the 112-fiber data (see 
Table~\ref{tabnum}). 

When there are more objects that meet the photometric
criteria than fibers available in a given 
1.5$\arcdeg$ x 1.5$\arcdeg$ field, the targets for spectra
 were chosen at random from within the photometric
boundaries.  Careful use of this ``galaxy sampling
fraction'' is required in subsequent analysis, but for the
112-fiber data that make up 80\% of the sample, the galaxy
selection fraction is large ($70\%$) and the details of this
procedure do not affect the derived results for quantities like 
the luminosity density, once the sampling is taken into account.
As described below, a small number of galaxy positions were
physically impossible to observe because of fiber ``collisions'',
and their effects on the statistical measures also need to be
assessed.  In the unusual case 
where there were fewer
galaxies than fibers for an individual field, the photometric limits
were enlarged a small amount to admit additional targets until all the 
fibers were in use. However,
for simplicity in the later data analyses, these additional galaxies, which 
constitute $< 10\%$ of the total data set, are not considered part of the 
survey proper. 

Table~\ref{tabdata} details the right ascension and declination borders, 
photometric limits $m_1$, $m_2$, and $m_{cen}$, number of galaxy redshifts
measured, and galaxy sampling fractions for each of the 327 
spectroscopic fields of
the Las Campanas survey proper. There
are small variations in $m_1$, $m_2$, and $m_{cen}$ from the nominal values
given above which result from recalibrations 
of the photometric zeropoints for the fields after the galaxies were
selected and observed.   Additional variations
in $m_{cen}$ are caused by differences in pixel size 
of the 3 CCD's
used during the course of the survey, and by attempts to account for 
shifts in $m_c$ caused by variable seeing on the drift-scan images. 
Most photometry for galaxies observed with the 50-fiber
system was obtained using our original 800 x 800 TI CCD, with
1$\farcs$06 pixels. The quality of the images was much improved for the 
larger chips used to obtain the 112-fiber data photometry and the analysis of the
luminosity function (\cite{lin96a}) explicitly separates the 50-fiber data 
from the 112-fiber data to compute separate selection functions and
luminosity functions for the two sets of data.

\subsection{Spectroscopy and Redshift Measurements} \label{spec}

A detailed description of the Las Campanas fiber system is given by
Shectman (1993). The silica fibers have a 3$\farcs$5 diameter in the focal
plane, are safely encased in concentric layers of hypodermic needle tubing
and bicycle brake cable housing, and are
manually plugged into holes on a 90-cm aluminum plate mounted
at the curved focal plane of the Las Campanas 2.5-m DuPont telescope. 
The fibers guide the light from the objects into a
spectrograph on the observing floor where the spectra are detected by the 2D-Frutti 
two-dimensional photon counter (\cite{she84}).  This combination
consitutes the ``fruit and fiber'' system.
Each plate is pre-drilled by a computer-controlled milling machine
in Pasadena
with holes at the expected positions of the selected 
objects.  Holes for objects on four separate fields
were drilled
on a single plate, so that over 400 redshifts could be obtained during a
single night using the 112-fiber system, without changing plates.

Mechanical constraints prevent the fibers from being plugged closer
together than 55$''$.  We have 
maintained a list of objects (about 1100) which were not observed
because they were too close to another hole in the plate.  
In all other respects, these objects fulfill the selection criteria, 
including the random selection, and the list can be used to correct
the small-scale correlation function and other statistics of the
galaxy distribution.  The galaxies which were not
observed because of ``collisions'' are flagged in Table 3, the detailed
listing of redshifts available on CD-ROM. 
Each field is one half of a photometric brick, with dimensions
$1.5 \arcdeg \times 1.5 \arcdeg$, and the exposure times are typically
2 hours per field. Under 30 minutes are needed to change the 
observing setup from one exposure to the next, so a typical clear night produces
four fields.
For the 50-fiber system, there were an additional 
10 fibers dedicated to observing sky spectra, and for the 112-fiber
system, 16 sky fibers.  Because the galaxies in the
LCRS are equal to or brighter than the sky as observed through a fiber,
precise sky subtraction was not required, and this sample of 
sky spectra proved adequate for our purposes.

The spectra are extracted, sky-subtracted, and wavelength-calibrated
using a combination of custom programs
and standard IRAF routines. For sky subtraction, all the sky spectra of a 
given exposure are averaged, then that average sky spectrum is normalized 
to the bright [OI] 5577 \AA \ sky line of each raw object+sky spectrum,
to account for individual fiber transmissions, before being
subtracted. For 
wavelength calibration we use He-Ne arc lamp spectra taken with the fiber
system midway through each object exposure. Some 35 lines in the range
3000-7000~\AA \ are used to derive the wavelength solution. The typical rms
wavelength residual for the comparison lines is 0.7~\AA . The spectra are
finally linearized and rebinned, at 2.5~\AA \ per pixel (approximating the 
original pixel size), to the wavelength range 3350-6400~\AA \ for the 
50-fiber data or 3350-6750~\AA \ for the 112-fiber data.

\begin{figure}
\plotone{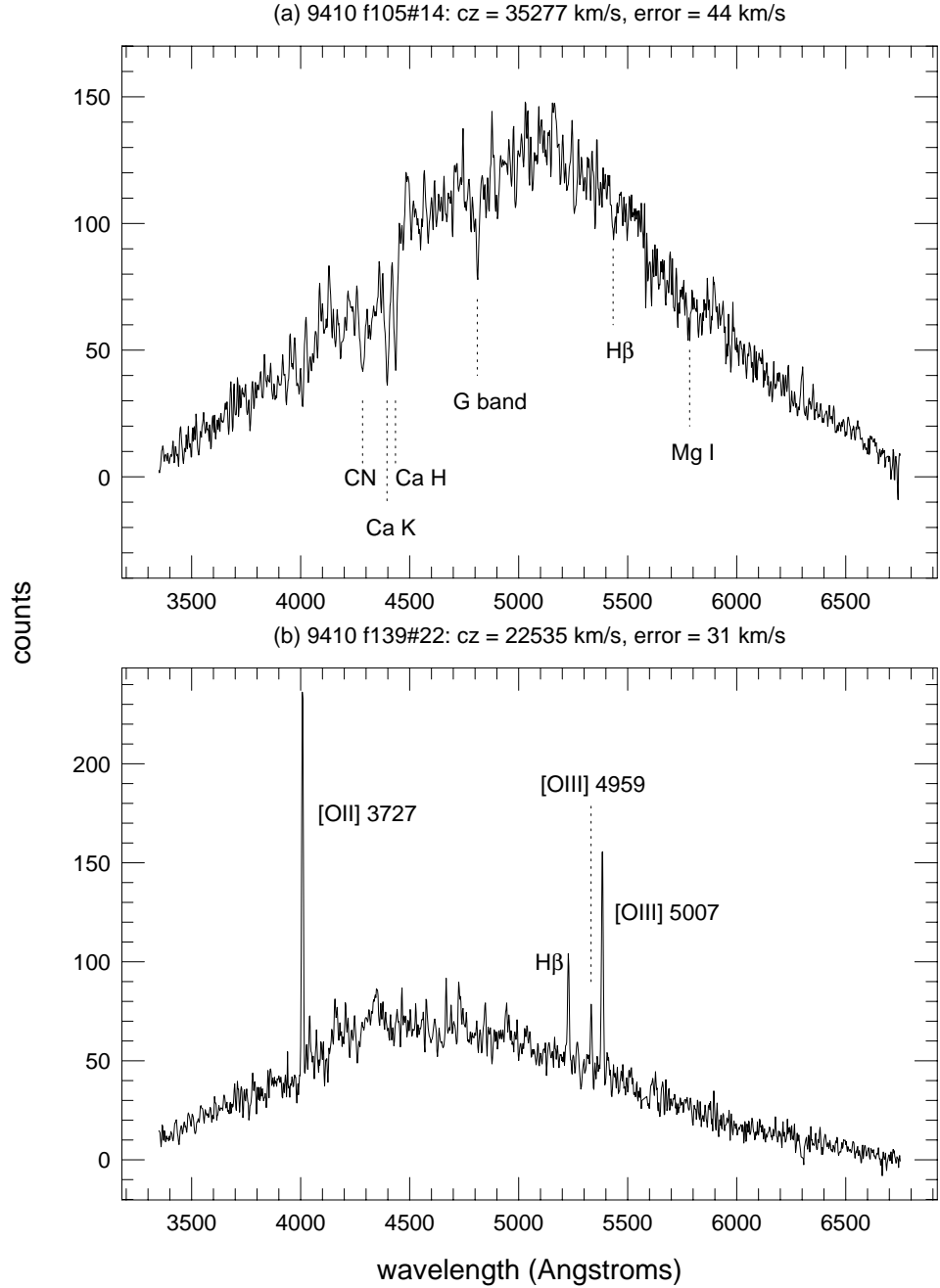}
\caption{Two example LCRS galaxy spectra. (a) A galaxy with prominent 
   absorption features. (b) A galaxy showing strong 
   emission lines.}
\label{figspec}
\end{figure}

Redshifts are determined by cross-correlating the object spectra 
against a set of template spectra.
The standard cross-correlation technique (\cite{ton79}) is implemented
using the IRAF add-on package RVSAO (\cite{kur91}, \cite{min95}). 
We use three templates, each consisting of
an average of many high signal-to-noise stellar spectra. Two of the templates
have strong absorption features typical of non-emission-line galaxies
(see \cite{san75}): CN (3833~\AA), Ca~H (3969~\AA) and K (3934~\AA), 
G-band (4304~\AA), and Mg I (5175~\AA). A third template contains strong
Balmer absorption features: H$\beta$ (4861~\AA), H$\gamma$ (4340~\AA), and 
H$\delta$ (4102~\AA). The third template is 
useful for obtaining cross-correlation velocities for 
the absorption-line features in galaxies with emission lines, 
for which such early-type Balmer absorption 
features are often seen. The third template
avoids a systematic bias that creeps into cross-correlation velocities 
for the emission-line galaxies. This problem occurs for
the first two templates because of systematic differences in the 
aborption-line spectra
 between the two 
templates and the typical galaxy with emission lines, particularly the increasing
blend of H$\epsilon$ 3970~\AA \ with Ca~H.
Gaussian profiles
are fit to emission lines to determine redshifts from the line
centers and also to measure line equivalent widths. The five most prominent
emission lines present in our spectra are: [OII] 3727, [OIII] 5007, [OIII] 4959,
H$\beta$ and H$\gamma$. The emission velocities from these lines
are consistent to about 5~km~s$^{-1}$, providing a check on our
wavelength solutions, and
the emission and third-template cross-correlation velocity zeropoints
agree to 15~km~s$^{-1}$.
Two sample spectra, one with prominent absorption
features and one with strong emission lines, are shown 
in Figure~\ref{figspec}.  One of the distinguishing features of the
LCRS is that it has produced a very large and homogeneous set of
galaxy spectra which are useful for characterizing the stellar
populations and gas content of galaxies, not just their velocities 
(e.g., \cite{zab96}).

Every spectrum is visually inspected at least once to check the plausibility of
the automated cross-correlation and emission velocity determinations.
Poor signal-to-noise spectra (approximately the bottom third) are 
examined more carefully
and the final velocity determination, or rejection of the spectrum as
a failure, is made interactively. We find that 93\% of our spectra are
galaxies, another 3\% are stars contaminating our sample, and the
final 4\% fail to yield either a galaxy redshift or an identification 
as a star. Our rate for identifying
spectra as galaxies or stars does not vary strongly with isophotal 
magnitude or central surface brightness; the identification rate only
drops by about 5\% at the faint isophotal magnitude limit or at
the central surface brightness cut line (\cite{lin96a}).
Our failure rate increases toward the edges of the DuPont telescope
field (Shectman et al. 1995).  This is a subtle effect whose impact on
clustering statistics should be small, but which we are evaluating.
Of the galaxy redshifts, 68\% are cross-correlation velocities
based on absorption lines alone, 
25\% are a combination of absorption and emission velocities, and 
the remaining 7\% are solely emission velocities.
A total of 26418 galaxy
redshifts have been obtained in the course of the survey; 23697 of 
these galaxies
lie within the photometric limits and geometric 
boundaries of the survey proper. 
The mean galaxy sampling fraction, corrected for stellar contamination,
is 70\% for the 112-fiber data and 58\% for the 50-fiber data.
Table~\ref{tabnum} summarizes some of the above numbers for the main 
survey data samples.
Because the
spectroscopic fields are not observed more than once, the sampling fractions
must be tracked on a field-by-field basis (as in Table~\ref{tabdata}) 
in subsequent statistical analyses of the survey data.  Other
surveys such as the Century Survey or the ESO Survey return to fields to 
observe every galaxy that meets
their selection criterion, so it will be informative to compare
results from these surveys with the LCRS to assess the effects of this
choice. We expect they will be small (\cite{lin96b}; \cite{tuc96xi}), 
but they are in the sense of the
LCRS undersampling fields which have an unusually high density of
galaxies, such as galaxy clusters.

\begin{figure}
\plotone{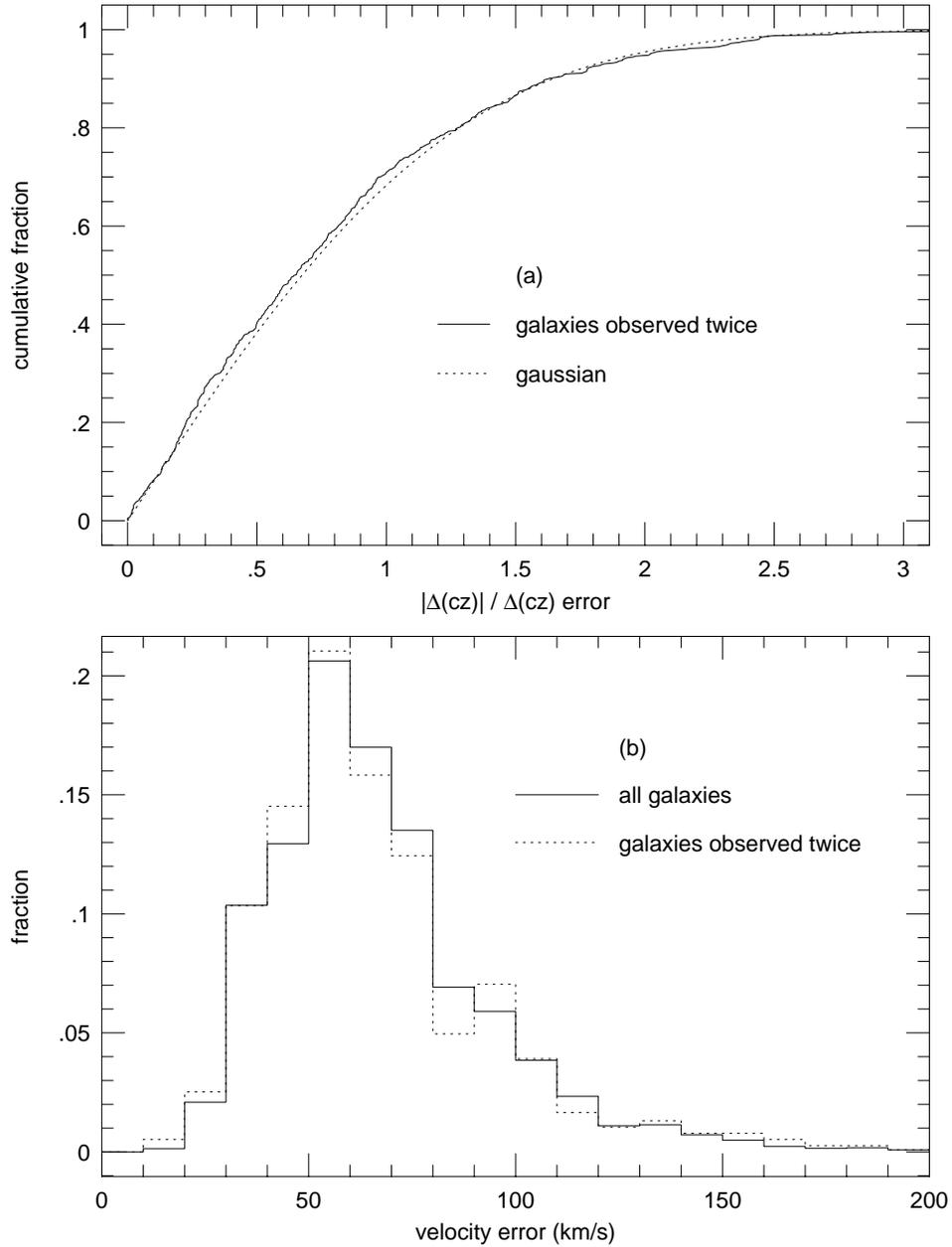}
\caption{(a) The cumulative distribution of velocity differences for 
   LCRS galaxies that were observed twice. The velocity
   differences have been normalized by their errors, and the
   distribution is consistent with a gaussian of unit variance.
   (b) The distribution of velocity errors for all LCRS galaxies
   and for LCRS galaxies which were observed twice.}
\label{figec}
\end{figure}

The cross-correlation and 
emission-line
fitting IRAF programs generate formal errors. We checked these estimates using a sample of
575 galaxies which were successfully observed twice, in the small overlapping
areas between some of our spectroscopic fields. For our data, the formal errors
underestimate the true random errors by 30\%, so we multiply the formal
errors by 1.3. Otherwise, application of a
Kolmogorov-Smirnov (K-S) test (e.g. \cite{pre92}, \S~14.3) shows that for
these repeated galaxies, the shape of the cumulative distribution of velocity differences,
normalized by the (corrected) velocity errors, is consistent with a
gaussian distribution; see Figure~\ref{figec}(a). Figure~\ref{figec}(b) 
shows the distribution of corrected velocity errors for all galaxies and for
the repeated galaxies; note that the two distributions are similar, suggesting that the
result derived for the overlap sample applies equally well to the whole survey.
The average random velocity error is 67~km~s$^{-1}$, corresponding to
$< 1 \ h^{-1}$~Mpc blurring
in the spatial domain, so that none of the effects described by Schuecker, Ott and Seitter
(1994) for low velocity precision redshifts is a problem for the LCRS.
A measure of the zeropoint offset in the 
velocities is provided by the velocities of the stellar spectra,
which should be near zero in the mean; we find $10 \pm 1$~km~s$^{-1}$ for
the average of 1102 stars, not exactly zero but comfortably smaller 
than our random 
velocity errors.
 
\section{The Survey Data} \label{data}

\begin{figure}
\plotfiddle{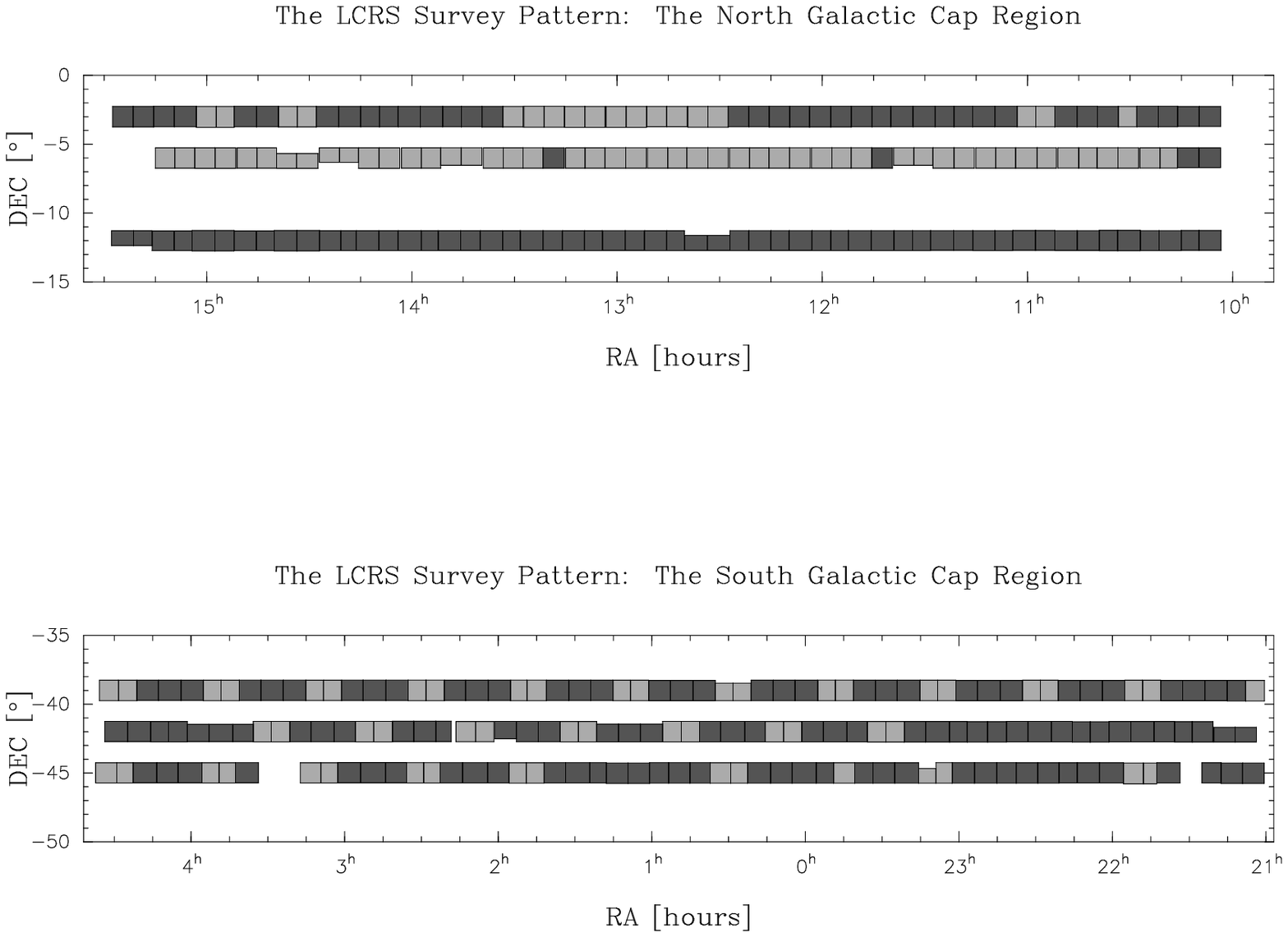}{17cm}{0}{90}{90}{-275}{-250}
\caption{The survey pattern for the North (top) and the South Galactic
   Cap regions (bottom) sampled by the LCRS.  Lightly shaded regions
   denote fields observed with the 50-fiber spectrograph and 
   darkly shaded regions fields observed with the 112-fiber 
   spectrograph. Declination and right ascension coordinates are
   epoch 1950.0.}
\label{figfields}
\end{figure}

The geometry of the survey is that of six ``slices'', each about $1.5 \arcdeg$
in declination by $80 \arcdeg$ in right ascension. Three slices are located
in the North galactic cap, centered at declinations $\delta = -3\arcdeg$, 
$-6\arcdeg$, and $-12\arcdeg$, and ranging in right ascension from 
$\alpha = 10^h$ to $15.5^h$. The other three slices are
in the South galactic cap, centered at declinations $\delta = -39\arcdeg$, 
$-42\arcdeg$, and $-45\arcdeg$, and ranging in right ascension from 
$\alpha = 21^h$ to $4.5^h$. The fields are at galactic
latitudes $b > 30\arcdeg$ in the North and $b < -40\arcdeg$ in the South.
Figure~\ref{figfields} shows the pattern of survey fields in
declination and right ascension on the sky, with different shadings
for 50- and 112-fiber fields. (Note that three fields in the 
$-45\arcdeg$ slice were not finished, thus producing gaps in the that
slice. There are also other small gaps between fields that cannot be
seen from the figures, so that one should always consult Table~\ref{tabdata}
for the exact boundaries.)
The distributions of LCRS galaxies in redshift space, as a function of
heliocentric velocity and 
right ascension, as well as their distributions on the sky, 
are next shown in Figures~\ref{fig-3}~(a-f)
for each of the six slices, and in Figures~\ref{fig-3}~(g,h) for
the combined Northern and Southern samples.
For the $\delta = -3\arcdeg$ slice and the three Southern
slices, the data combines 50-fiber and 112-fiber observations. The
factor of two difference in sampling can be seen in the lower density 
of points for the 50-fiber fields in the figures. The $\delta = -6\arcdeg$ 
slice is nearly all 50-fiber data, and the
$\delta = -12\arcdeg$ slice consists only of 112-fiber fields.

The figures demonstrate the rich texture of clusters, filaments, voids, and 
walls in the LCRS galaxy distribution, reminiscent of structures
seen in the denser, wider, and shallower CfA survey 
(which has a mean distance of about 7500~km~s$^{-1}$), 
but replicated many times across the bigger LCRS
volume. The largest walls and voids have sizes of order 
$50-100 \ h^{-1}$~Mpc, much smaller than the largest survey dimensions,
suggesting that the LCRS samples the largest high-contrast structures of 
the nearby
universe.  The 3$\arcdeg $ separation between slices amounts to about
$15 \ h^{-1}$~Mpc at a typical survey redshift of 30000 km~s$^{-1}$.  
The strong resemblance of one strip
to its neighbor and the structures that are visible in the combined
North or South fields provide another indication that coherent
structures on scales of tens of Mpc are very common. The redshift data
are found in Table 3, which is available in the AAS CD-ROM series.

Finally, redshift histograms for the 50- and 112-fiber North and South data 
sets are shown in Figures~\ref{figh50} and \ref{figh112} along with the expected 
redshift distributions computed for galaxies uniformly 
distributed in space and sampled using the survey's selection function 
as derived by Lin et al. (1996a). The average distance of 
the observed distribution is about $cz = 30000$~km~s$^{-1}$ as intended. 
Also, although large-scale structure makes the histogram noisy, the uniform 
distribution
line is a reasonable approximation to the actual redshift distribution,
which supports the view 
that the LCRS is large enough to approximate a fair sample of the
nearby Universe.

We welcome other investigators to apply their analysis techniques to this
large sample of galaxy redshifts.  For this reason, the redshift
catalog will be available at the LCRS home page 
(``http://manaslu.astro.utoronto.ca/\~{ }lin/lcrs.html'')
and in the AAS CD-ROM series.  To obtain reliable scientific results
from this redshift data, users need to be 
aware of the details of our selection methods described and tabulated in this paper.
These include the limits on isophotal magnitude and on central surface brightness,
the differences between data taken with the 50 fiber system and with the 112 fiber
system, and the limit on closest approach of the fibers.  For many purposes,
these effects can be taken into account quite easily, but the best use of
the survey requires attention to these details.


\acknowledgments
We thank the Yale Service Observers for
obtaining the photometric calibration frames at the CTIO 0.9 meter and
Prof. Suzanne W. Tourtellotte of Albertus Magnus College for her help
in performing many of the calibration frame reductions.
Thanks also to Paulo S. Pellegrini for a helpful referee's report.
The Las Campanas Redshift Survey has been supported
by NSF grants AST 87-17207, AST 89-21326, and AST 92-20460. HL also
acknowledges support from NASA grant NGT-51093.

\clearpage

\begin{deluxetable}{lrrrrrrrrrr}

\tablewidth{0pt}

\tablecaption{LCRS Sample Information \label{tabnum}}

\tablehead{
\colhead{Sample} &
\colhead{$N_{sel}$ \tablenotemark{a}} &
\colhead{$N_{lsb}$  \tablenotemark{a}} & \colhead{\%} &
\colhead{$N_{gal}$  \tablenotemark{b}} & \colhead{\%} &
\colhead{$N_{star}$ \tablenotemark{b}} & \colhead{\%} &
\colhead{$N_{?}$    \tablenotemark{b}} & \colhead{\%} &
\colhead{$f$ \tablenotemark{c}}
}
\startdata
North 50  & 4288 & 1179 & 22
          & 2711 & 92.2 & 130 & 4.4 &  99 & 3.4 & 0.66 \nl
South 50  & 4363 & 1145 & 21 
          & 2055 & 91.1 & 126 & 5.6 &  74 & 3.3 & 0.50 \nl
All 50    & 8651 & 2324 & 21
          & 4766 & 91.7 & 256 & 4.9 & 173 & 3.3 & 0.58 \nl
& & & & & & & & & & \nl
North 112 & 12220 & 557 &  4
          &  8552 & 94.9 & 153 & 1.7 & 302 & 3.4 & 0.71 \nl
South 112 & 15639 & 1530 & 9
          & 10379 & 91.4 & 451 & 4.0 & 525 & 4.6 & 0.69 \nl
All 112   & 27859 & 2087 & 7
          & 18931 & 93.0 & 604 & 3.0 & 827 & 4.1 & 0.70 \nl
\enddata
\tablenotetext{a}{$N_{sel}$ is the number of objects which lie within the
                  photometric limits and geometric 
                  borders of the sample. $N_{lsb}$
                  is the number of objects which were excluded because of
                  low central surface brightness (lsb) but were otherwise 
                  within the isophotal magnitude limits. The percentage denotes
                  the proportion of lsb objects among all objects within the
                  isophotal magnitude limits.}
\tablenotetext{b}{$N_{gal}$, $N_{star}$, and $N_{?}$ denote the number of
                  galaxy, stellar, and unidentified spectra, respectively,
                  which were obtained for those objects that met
                  the photometric and boundary limits of the sample. The
                  percentages refer to the proportions of each of the
                  three types among the spectra which were obtained.}
\tablenotetext{c}{$f$ is the galaxy sampling fraction for each sample,
                  corrected for stellar contamination. $f$ is estimated
                  by $f = (N_{gal} + N_{star}) / N_{sel}$. We thus assume
                  that the proportion of galaxies and stars among objects
                  with identified spectra is the same as that among
                  objects which were not observed or whose spectra failed
                  to yield an identification.}
\end{deluxetable}

\begin{deluxetable}{lcccccccrrc}
\footnotesize
\tablecaption{LCRS Spectroscopic Fields \label{tabdata}}
\tablewidth{0pt}
\tablehead{
\colhead{Field} &
\colhead{$\alpha_1$  \tablenotemark{a}} & 
\colhead{$\alpha_2$  \tablenotemark{a}} & 
\colhead{$\delta_1$  \tablenotemark{a}} & 
\colhead{$\delta_2$  \tablenotemark{a}} & 
\colhead{$m_1$  \tablenotemark{b}} & 
\colhead{$m_2$  \tablenotemark{b}} & 
\colhead{$m_{cen}$ \tablenotemark{b}} & 
\colhead{$N_{fib}$ \tablenotemark{c}} & 
\colhead{$N_{gal}$ \tablenotemark{d}} &
\colhead{$f$ \tablenotemark{e}} 
}
\startdata 
 1003-03W &  10 03 30.60 &  10 09 46.42 & -03 42 28.1  & -02 16 03.4  &
 15.06 & 17.76 & 19.41 & 112 & 111 & 0.87 \nl
 1003-03E &  10 09 46.42 &  10 16 05.81 & -03 42 28.1  & -02 16 03.4  &
 15.06 & 17.76 & 19.41 & 112 & 109 & 0.75 \nl
 1015-03W &  10 16 05.81 &  10 21 46.37 & -03 44 11.8  & -02 15 11.2  &
 14.86 & 17.56 & 18.71 & 112 &  93 & 0.56 \nl
 1015-03E &  10 21 46.37 &  10 28 00.72 & -03 44 11.8  & -02 15 11.2  &
 14.86 & 17.56 & 18.71 & 112 & 109 & 0.57 \nl
 1027-03W &  10 28 00.72 &  10 33 26.78 & -03 44 10.7  & -02 15 22.0  &
 16.01 & 17.31 & 18.31 &  50 &  42 & 0.58 \nl
 1027-03E &  10 33 26.78 &  10 39 33.77 & -03 44 10.7  & -02 15 22.0  &
 14.91 & 17.61 & 18.76 & 112 & 106 & 0.65 \nl
 1039-03W &  10 39 33.77 &  10 45 30.41 & -03 44 16.8  & -02 15 14.4  &
 14.94 & 17.64 & 18.79 & 112 &  97 & 0.67 \nl
 1039-03E &  10 45 30.41 &  10 52 00.53 & -03 44 16.8  & -02 15 14.4  &
 14.94 & 17.64 & 18.79 & 112 & 101 & 0.79 \nl
 1051-03W &  10 52 00.53 &  10 57 34.87 & -03 44 11.8  & -02 15 09.4  &
 16.09 & 17.39 & 18.29 &  50 &  32 & 0.89 \nl
 1051-03E &  10 57 34.87 &  11 03 00.26 & -03 44 11.8  & -02 15 09.4  &
 16.09 & 17.39 & 18.29 &  50 &  33 & 0.94 \nl
 1103-03W &  11 03 00.26 &  11 09 30.94 & -03 43 53.0  & -02 14 52.4  &
 14.94 & 17.64 & 18.79 & 112 & 105 & 0.55 \nl
 1103-03E &  11 09 30.94 &  11 16 00.74 & -03 43 53.0  & -02 14 52.4  &
 14.94 & 17.64 & 18.79 & 112 & 109 & 0.54 \nl
 1115-03W &  11 16 00.74 &  11 21 29.57 & -03 44 03.5  & -02 15 02.2  &
 14.73 & 17.53 & 18.63 & 112 &  95 & 0.83 \nl
 1115-03E &  11 21 29.57 &  11 27 00.17 & -03 44 03.5  & -02 15 02.2  &
 14.73 & 17.53 & 18.63 & 112 &  67 & 0.85 \nl
 1127-03W &  11 27 00.17 &  11 33 28.37 & -03 44 10.7  & -02 14 55.7  &
 14.88 & 17.58 & 18.73 & 112 & 109 & 0.71 \nl
 1127-03E &  11 33 28.37 &  11 39 58.06 & -03 44 10.7  & -02 14 55.7  &
 14.88 & 17.58 & 18.73 & 112 & 105 & 0.52 \nl
 1139-03W &  11 39 58.06 &  11 45 26.45 & -03 44 10.3  & -02 15 05.8  &
 14.81 & 17.61 & 18.71 & 112 &  97 & 0.57 \nl
 1139-03E &  11 45 26.45 &  11 51 31.18 & -03 44 10.3  & -02 15 05.8  &
 14.81 & 17.61 & 18.71 & 112 & 110 & 0.43 \nl
 1151-03W &  11 51 31.18 &  11 57 30.24 & -03 44 33.7  & -02 15 08.6  &
 15.01 & 17.71 & 18.86 & 112 &  88 & 0.86 \nl
 1151-03E &  11 57 30.24 &  12 03 42.31 & -03 44 33.7  & -02 15 08.6  &
 15.01 & 17.71 & 18.86 & 112 &  95 & 0.90 \nl
 1203-03W &  12 03 42.31 &  12 09 31.32 & -03 44 41.3  & -02 15 16.2  &
 15.05 & 17.65 & 18.85 & 112 &  96 & 0.89 \nl
 1203-03E &  12 09 31.32 &  12 15 31.30 & -03 44 41.3  & -02 15 16.2  &
 15.05 & 17.65 & 18.85 & 112 & 108 & 0.59 \nl
 1215-03W &  12 15 31.30 &  12 21 28.97 & -03 44 12.5  & -02 15 09.7  &
 14.79 & 17.59 & 18.69 & 112 & 104 & 0.86 \nl
 1215-03E &  12 21 28.97 &  12 27 29.74 & -03 44 12.5  & -02 15 09.7  &
 14.79 & 17.59 & 18.69 & 112 & 104 & 0.89 \nl
 1227-03W &  12 27 29.74 &  12 33 27.70 & -03 46 05.2  & -02 15 20.9  &
 15.92 & 17.22 & 18.07 &  50 &  33 & 0.82 \nl
 1227-03E &  12 33 27.70 &  12 39 18.65 & -03 46 05.2  & -02 15 20.9  &
 15.92 & 17.22 & 18.07 &  50 &  48 & 0.91 \nl
 1239-03W &  12 39 28.66 &  12 45 27.94 & -03 43 28.9  & -02 15 25.9  &
 15.85 & 17.15 & 18.00 &  50 &  25 & 0.45 \nl
 1239-03E &  12 45 27.94 &  12 51 19.46 & -03 43 28.9  & -02 15 25.9  &
 15.85 & 17.15 & 18.00 &  50 &  43 & 0.76 \nl
 1251-03W &  12 51 23.11 &  12 57 21.41 & -03 46 22.4  & -02 15 34.9  &
 15.83 & 17.13 & 18.08 &  50 &  47 & 0.63 \nl
 1251-03E &  12 57 21.41 &  13 03 13.54 & -03 46 22.4  & -02 15 34.9  &
 15.83 & 17.13 & 18.08 &  50 &  47 & 0.37 \nl
 1303-03W &  13 03 30.26 &  13 09 25.20 & -03 46 21.0  & -02 15 19.4  &
 16.00 & 17.30 & 18.15 &  50 &  45 & 0.68 \nl
 1303-03E &  13 09 25.20 &  13 15 19.58 & -03 46 21.0  & -02 15 19.4  &
 16.00 & 17.30 & 18.15 &  50 &  39 & 0.89 \nl
 1315-03W &  13 15 26.62 &  13 21 20.71 & -03 45 55.1  & -02 15 04.3  &
 15.88 & 17.18 & 18.13 &  50 &  49 & 0.67 \nl
 1315-03E &  13 21 20.71 &  13 27 13.63 & -03 45 55.1  & -02 15 04.3  &
 15.88 & 17.18 & 18.13 &  50 &  50 & 0.53 \nl
 1327-03W &  13 27 28.51 &  13 33 24.96 & -03 44 44.2  & -02 15 39.2  &
 16.03 & 17.33 & 18.33 &  50 &  50 & 0.59 \nl
 1327-03E &  13 33 24.96 &  13 39 29.93 & -03 44 44.2  & -02 15 39.2  &
 14.93 & 17.63 & 18.78 & 112 & 107 & 0.81 \nl
 1339-03W &  13 39 30.62 &  13 45 30.00 & -03 44 20.4  & -02 15 33.8  &
 14.81 & 17.61 & 18.71 & 112 &  78 & 0.72 \nl
 1339-03E &  13 45 30.00 &  13 50 59.86 & -03 44 20.4  & -02 15 33.8  &
 14.81 & 17.61 & 18.71 & 112 &  80 & 0.89 \nl
 1351-03W &  13 50 59.86 &  13 57 31.73 & -03 44 24.7  & -02 15 29.9  &
 14.77 & 17.47 & 18.62 & 112 & 104 & 0.93 \nl
 1351-03E &  13 57 31.73 &  14 04 00.00 & -03 44 24.7  & -02 15 29.9  &
 14.77 & 17.47 & 18.62 & 112 &  96 & 0.86 \nl
 1403-03W &  14 04 00.00 &  14 09 30.10 & -03 44 11.4  & -02 15 21.2  &
 14.78 & 17.58 & 18.68 & 112 &  94 & 0.85 \nl
 1403-03E &  14 09 30.10 &  14 14 59.52 & -03 44 11.4  & -02 15 21.2  &
 14.78 & 17.58 & 18.68 & 112 &  89 & 0.92 \nl
 1415-03W &  14 14 59.52 &  14 21 24.12 & -03 44 21.1  & -02 15 18.7  &
 15.01 & 17.71 & 18.86 & 112 & 103 & 0.91 \nl
 1415-03E &  14 21 24.12 &  14 27 57.67 & -03 44 21.1  & -02 15 18.7  &
 15.01 & 17.71 & 18.86 & 112 & 108 & 0.91 \nl
 1427-03W &  14 27 57.67 &  14 33 31.15 & -03 44 47.8  & -02 15 49.3  &
 16.14 & 17.44 & 18.44 &  50 &  26 & 0.83 \nl
 1427-03E &  14 33 31.15 &  14 39 00.34 & -03 44 47.8  & -02 15 49.3  &
 16.14 & 17.44 & 18.44 &  50 &  41 & 0.88 \nl
 1439-03W &  14 39 00.34 &  14 45 25.46 & -03 44 17.5  & -02 15 14.0  &
 15.11 & 17.81 & 18.96 & 112 & 107 & 0.72 \nl
 1439-03E &  14 45 25.46 &  14 51 57.19 & -03 44 17.5  & -02 15 14.0  &
 15.11 & 17.81 & 18.96 & 112 & 108 & 0.91 \nl
 1451-03W &  14 51 57.19 &  14 57 11.54 & -03 46 22.8  & -02 15 37.1  &
 15.96 & 17.26 & 18.11 &  50 &  37 & 0.70 \nl
 1451-03E &  14 57 11.54 &  15 03 02.04 & -03 46 22.8  & -02 15 37.1  &
 15.96 & 17.26 & 18.11 &  50 &  25 & 0.97 \nl
 1503-03W &  15 03 02.04 &  15 09 30.22 & -03 44 12.1  & -02 15 14.0  &
 14.76 & 17.56 & 18.66 & 112 &  63 & 0.92 \nl
 1503-03E &  15 09 30.22 &  15 15 30.17 & -03 44 12.1  & -02 15 14.0  &
 14.76 & 17.56 & 18.66 & 112 & 100 & 0.72 \nl
 1515-03W &  15 15 30.17 &  15 21 30.82 & -03 44 38.0  & -02 15 43.6  &
 14.89 & 17.69 & 18.79 & 112 &  66 & 0.91 \nl
 1515-03E &  15 21 30.82 &  15 27 30.02 & -03 44 38.0  & -02 15 43.6  &
 14.89 & 17.69 & 18.74 & 112 &  62 & 0.91 \nl
 1003-06W &  10 03 30.96 &  10 09 52.32 & -06 42 12.6  & -05 15 49.0  &
 15.00 & 17.70 & 19.35 & 112 & 100 & 0.55 \nl
 1003-06E &  10 09 52.32 &  10 16 08.62 & -06 42 12.6  & -05 15 49.0  &
 15.00 & 17.70 & 19.35 & 112 &  92 & 0.88 \nl
 1015-06W &  10 16 08.62 &  10 21 17.71 & -06 43 56.3  & -05 15 33.8  &
 15.98 & 17.28 & 18.13 &  50 &  38 & 0.39 \nl
 1015-06E &  10 21 17.71 &  10 27 06.96 & -06 43 56.3  & -05 15 33.8  &
 15.98 & 17.28 & 18.13 &  50 &  44 & 0.81 \nl
 1027-06W &  10 27 30.82 &  10 33 33.17 & -06 44 26.9  & -05 15 13.3  &
 15.92 & 17.22 & 18.22 &  50 &  45 & 0.66 \nl
 1027-06E &  10 33 33.17 &  10 39 37.18 & -06 44 26.9  & -05 15 13.3  &
 15.92 & 17.22 & 18.22 &  50 &  38 & 0.76 \nl
 1039-06W &  10 39 37.18 &  10 45 02.45 & -06 43 53.4  & -05 15 25.2  &
 16.26 & 17.56 & 18.41 &  50 &  44 & 0.90 \nl
 1039-06E &  10 45 02.45 &  10 51 06.14 & -06 43 53.4  & -05 15 25.2  &
 16.26 & 17.56 & 18.41 &  50 &  46 & 0.48 \nl
 1051-06W &  10 51 31.66 &  10 57 11.11 & -06 44 15.0  & -05 15 23.0  &
 16.08 & 17.38 & 18.28 &  50 &  48 & 0.60 \nl
 1051-06E &  10 57 11.11 &  11 03 27.31 & -06 44 15.0  & -05 15 23.0  &
 16.08 & 17.38 & 18.28 &  50 &  48 & 0.66 \nl
 1103-06W &  11 03 28.44 &  11 09 18.22 & -06 44 19.0  & -05 15 27.4  &
 16.03 & 17.33 & 18.28 &  50 &  40 & 0.93 \nl
 1103-06E &  11 09 18.22 &  11 15 05.64 & -06 44 19.0  & -05 15 27.4  &
 16.03 & 17.33 & 18.28 &  50 &  39 & 0.51 \nl
 1115-06W &  11 15 30.91 &  11 21 28.03 & -06 44 31.6  & -05 15 26.3  &
 16.08 & 17.38 & 18.38 &  50 &  48 & 0.83 \nl
 1115-06E &  11 21 28.03 &  11 27 38.09 & -06 44 31.6  & -05 15 26.3  &
 16.08 & 17.38 & 18.38 &  50 &  50 & 0.85 \nl
 1127-06W &  11 27 38.09 &  11 33 11.81 & -06 31 41.2  & -05 15 32.0  &
 16.01 & 17.31 & 18.16 &  50 &  48 & 0.69 \nl
 1127-06E &  11 33 11.81 &  11 39 05.69 & -06 31 41.2  & -05 15 32.0  &
 16.01 & 17.31 & 18.16 &  50 &  36 & 0.93 \nl
 1139-06W &  11 39 28.82 &  11 45 30.91 & -06 44 22.9  & -05 15 01.1  &
 14.98 & 17.68 & 18.83 & 112 & 109 & 0.74 \nl
 1139-06E &  11 45 30.91 &  11 51 34.61 & -06 44 22.9  & -05 15 01.1  &
 16.08 & 17.38 & 18.38 &  50 &  47 & 0.52 \nl
 1151-06W &  11 51 34.61 &  11 57 09.77 & -06 44 21.1  & -05 15 27.0  &
 16.04 & 17.34 & 18.19 &  50 &  45 & 0.65 \nl
 1151-06E &  11 57 09.77 &  12 03 07.08 & -06 44 21.1  & -05 15 27.0  &
 16.04 & 17.34 & 18.19 &  50 &  46 & 0.58 \nl
 1203-06W &  12 03 30.98 &  12 09 36.02 & -06 44 44.9  & -05 15 18.7  &
 16.02 & 17.32 & 18.22 &  50 &  46 & 0.63 \nl
 1203-06E &  12 09 36.02 &  12 15 36.74 & -06 44 44.9  & -05 15 18.7  &
 16.02 & 17.32 & 18.22 &  50 &  46 & 0.42 \nl
 1215-06W &  12 15 36.74 &  12 21 16.92 & -06 44 15.7  & -05 15 38.9  &
 16.07 & 17.37 & 18.22 &  50 &  41 & 0.92 \nl
 1215-06E &  12 21 16.92 &  12 27 07.68 & -06 44 15.7  & -05 15 38.9  &
 16.07 & 17.37 & 18.22 &  50 &  37 & 0.93 \nl
 1227-06W &  12 27 30.17 &  12 33 27.58 & -06 44 22.9  & -05 15 23.8  &
 16.16 & 17.46 & 18.36 &  50 &  38 & 0.76 \nl
 1227-06E &  12 33 27.58 &  12 39 36.74 & -06 44 22.9  & -05 15 23.8  &
 16.16 & 17.46 & 18.36 &  50 &  48 & 0.68 \nl
 1239-06W &  12 39 36.74 &  12 45 11.78 & -06 44 15.4  & -05 15 28.4  &
 16.13 & 17.43 & 18.28 &  50 &  43 & 0.47 \nl
 1239-06E &  12 45 11.78 &  12 51 06.84 & -06 44 15.4  & -05 15 28.4  &
 16.13 & 17.43 & 18.28 &  50 &  43 & 0.47 \nl
 1251-06W &  12 51 29.11 &  12 57 28.78 & -06 44 39.1  & -05 15 12.2  &
 16.10 & 17.40 & 18.40 &  50 &  46 & 0.76 \nl
 1251-06E &  12 57 28.78 &  13 03 33.10 & -06 44 39.1  & -05 15 12.2  &
 16.10 & 17.40 & 18.40 &  50 &  47 & 0.53 \nl
 1303-06W &  13 03 33.10 &  13 09 22.85 & -06 44 19.0  & -05 15 36.7  &
 16.28 & 17.58 & 18.43 &  50 &  44 & 0.51 \nl
 1303-06E &  13 09 22.85 &  13 15 05.14 & -06 44 19.0  & -05 15 36.7  &
 16.28 & 17.58 & 18.43 &  50 &  39 & 0.51 \nl
 1315-06W &  13 15 28.56 &  13 21 38.35 & -06 44 37.7  & -05 15 29.5  &
 15.00 & 17.70 & 18.85 & 112 &  68 & 0.71 \nl
 1315-06E &  13 21 38.35 &  13 27 29.02 & -06 44 37.7  & -05 15 29.5  &
 16.10 & 17.40 & 18.30 &  50 &  28 & 0.84 \nl
 1327-06W &  13 27 29.02 &  13 33 17.66 & -06 44 25.4  & -05 15 50.4  &
 15.98 & 17.28 & 18.13 &  50 &  50 & 0.93 \nl
 1327-06E &  13 33 17.66 &  13 39 07.70 & -06 44 25.4  & -05 15 50.4  &
 15.98 & 17.28 & 18.13 &  50 &  48 & 0.59 \nl
 1339-06W &  13 39 31.06 &  13 45 34.39 & -06 31 58.4  & -05 15 06.1  &
 16.03 & 17.33 & 18.23 &  50 &  47 & 0.70 \nl
 1339-06E &  13 45 34.39 &  13 51 35.93 & -06 31 58.4  & -05 15 06.1  &
 16.03 & 17.33 & 18.23 &  50 &  40 & 0.89 \nl
 1351-06W &  13 51 35.93 &  13 57 15.53 & -06 44 16.8  & -05 15 37.4  &
 15.83 & 17.13 & 17.98 &  50 &  31 & 0.92 \nl
 1351-06E &  13 57 15.53 &  14 03 04.18 & -06 44 16.8  & -05 15 37.4  &
 15.83 & 17.13 & 17.98 &  50 &  49 & 0.72 \nl
 1403-06W &  14 03 31.75 &  14 09 34.92 & -06 44 52.4  & -05 15 46.8  &
 16.07 & 17.37 & 18.37 &  50 &  43 & 0.64 \nl
 1403-06E &  14 09 34.92 &  14 15 36.55 & -06 44 52.4  & -05 15 46.8  &
 16.07 & 17.37 & 18.37 &  50 &  48 & 0.83 \nl
 1415-06W &  14 15 36.55 &  14 21 07.15 & -06 19 08.8  & -05 15 51.5  &
 16.06 & 17.36 & 18.21 &  50 &  36 & 0.82 \nl
 1415-06E &  14 21 07.15 &  14 27 06.41 & -06 19 08.8  & -05 15 51.5  &
 16.06 & 17.36 & 18.21 &  50 &  17 & 0.86 \nl
 1427-06W &  14 27 28.70 &  14 33 45.29 & -06 44 51.4  & -05 40 55.2  &
 16.10 & 17.40 & 18.30 &  50 &  26 & 0.86 \nl
 1427-06E &  14 33 45.29 &  14 39 35.18 & -06 44 51.4  & -05 40 55.2  &
 16.10 & 17.40 & 18.30 &  50 &  26 & 0.76 \nl
 1439-06W &  14 39 35.18 &  14 45 18.74 & -06 44 25.8  & -05 15 51.1  &
 16.18 & 17.48 & 18.33 &  50 &  45 & 0.66 \nl
 1439-06E &  14 45 18.74 &  14 51 06.48 & -06 44 25.8  & -05 15 51.1  &
 16.18 & 17.48 & 18.33 &  50 &  42 & 0.62 \nl
 1451-06W &  14 51 29.76 &  14 57 34.42 & -06 44 50.6  & -05 15 35.6  &
 16.13 & 17.43 & 18.43 &  50 &  30 & 0.66 \nl
 1451-06E &  14 57 34.42 &  15 03 28.32 & -06 44 50.6  & -05 15 35.6  &
 16.13 & 17.43 & 18.43 &  50 &  33 & 0.75 \nl
 1503-06W &  15 03 28.32 &  15 09 19.63 & -06 44 28.7  & -05 15 45.4  &
 16.04 & 17.34 & 18.29 &  50 &  44 & 0.61 \nl
 1503-06E &  15 09 19.63 &  15 15 00.38 & -06 44 28.7  & -05 15 45.4  &
 16.04 & 17.34 & 18.29 &  50 &  47 & 0.77 \nl
 1003-12W &  10 03 26.02 &  10 09 50.81 & -12 42 16.6  & -11 15 49.7  &
 14.91 & 17.61 & 19.26 & 112 &  79 & 0.87 \nl
 1003-12E &  10 09 50.81 &  10 15 00.58 & -12 42 16.6  & -11 15 49.7  &
 14.91 & 17.61 & 19.26 & 112 &  64 & 0.85 \nl
 1015-12W &  10 15 00.58 &  10 21 30.48 & -12 42 59.8  & -11 16 29.6  &
 15.02 & 17.72 & 19.27 & 112 &  97 & 0.80 \nl
 1015-12E &  10 21 30.48 &  10 27 00.53 & -12 42 59.8  & -11 16 29.6  &
 15.02 & 17.72 & 19.27 & 112 & 102 & 0.85 \nl
 1027-12W &  10 27 00.53 &  10 33 38.28 & -12 43 58.8  & -11 15 01.4  &
 15.01 & 17.71 & 18.86 & 112 & 107 & 0.77 \nl
 1027-12E &  10 33 38.28 &  10 39 00.22 & -12 43 58.8  & -11 15 01.4  &
 15.01 & 17.71 & 18.86 & 112 &  87 & 0.67 \nl
 1039-12W &  10 39 00.22 &  10 45 28.01 & -12 42 41.0  & -11 16 11.6  &
 15.03 & 17.73 & 19.28 & 112 & 103 & 0.68 \nl
 1039-12E &  10 45 28.01 &  10 51 59.69 & -12 42 41.0  & -11 16 11.6  &
 15.03 & 17.73 & 19.28 & 112 & 111 & 0.73 \nl
 1051-12W &  10 51 59.69 &  10 57 37.80 & -12 42 06.8  & -11 15 36.7  &
 15.08 & 17.78 & 19.43 & 112 &  93 & 0.49 \nl
 1051-12E &  10 57 37.80 &  11 04 16.75 & -12 42 06.8  & -11 15 36.7  &
 15.08 & 17.78 & 19.43 & 112 & 105 & 0.79 \nl
 1103-12W &  11 04 16.75 &  11 09 30.36 & -12 43 00.8  & -11 16 36.5  &
 15.03 & 17.73 & 19.28 & 112 &  69 & 0.89 \nl
 1103-12E &  11 09 30.36 &  11 15 58.27 & -12 43 00.8  & -11 16 36.5  &
 15.03 & 17.73 & 19.28 & 112 & 109 & 0.59 \nl
 1115-12W &  11 15 58.27 &  11 21 35.30 & -12 42 11.5  & -11 16 03.0  &
 15.01 & 17.71 & 19.36 & 112 &  90 & 0.82 \nl
 1115-12E &  11 21 35.30 &  11 28 18.50 & -12 42 11.5  & -11 16 03.0  &
 15.01 & 17.71 & 19.36 & 112 & 109 & 0.81 \nl
 1127-12W &  11 28 18.50 &  11 33 37.63 & -12 43 14.2  & -11 16 28.6  &
 15.01 & 17.71 & 19.26 & 112 &  99 & 0.51 \nl
 1127-12E &  11 33 37.63 &  11 39 01.63 & -12 43 14.2  & -11 16 28.6  &
 15.01 & 17.71 & 19.26 & 112 & 102 & 0.45 \nl
 1139-12W &  11 39 01.63 &  11 45 41.33 & -12 43 01.6  & -11 16 20.3  &
 15.00 & 17.70 & 19.35 & 112 & 110 & 0.54 \nl
 1139-12E &  11 45 41.33 &  11 52 20.90 & -12 43 01.6  & -11 16 20.3  &
 15.00 & 17.70 & 19.35 & 112 & 109 & 0.68 \nl
 1151-12W &  11 52 20.90 &  11 57 29.23 & -12 42 58.7  & -11 16 17.8  &
 15.03 & 17.73 & 19.28 & 112 &  90 & 0.79 \nl
 1151-12E &  11 57 29.23 &  12 03 00.91 & -12 42 58.7  & -11 16 17.8  &
 15.03 & 17.73 & 19.28 & 112 &  54 & 0.77 \nl
 1203-12W &  12 03 00.91 &  12 09 39.19 & -12 42 55.8  & -11 16 20.6  &
 15.03 & 17.73 & 19.38 & 112 &  75 & 0.89 \nl
 1203-12E &  12 09 39.19 &  12 15 00.48 & -12 42 55.8  & -11 16 20.6  &
 15.03 & 17.73 & 19.38 & 112 &  84 & 0.86 \nl
 1215-12W &  12 15 00.48 &  12 21 27.07 & -12 43 14.2  & -11 16 34.3  &
 15.04 & 17.74 & 19.29 & 112 &  82 & 0.79 \nl
 1215-12E &  12 21 27.07 &  12 26 58.68 & -12 43 14.2  & -11 16 34.3  &
 15.04 & 17.74 & 19.29 & 112 &  72 & 0.87 \nl
 1227-12W &  12 26 58.68 &  12 33 37.20 & -12 42 40.3  & -11 37 09.1  &
 15.10 & 17.80 & 19.45 & 112 & 105 & 0.87 \nl
 1227-12E &  12 33 37.20 &  12 40 19.08 & -12 42 40.3  & -11 37 09.1  &
 15.10 & 17.80 & 19.45 & 112 &  99 & 0.89 \nl
 1239-12W &  12 40 19.08 &  12 45 32.21 & -12 43 21.7  & -11 16 50.5  &
 14.92 & 17.62 & 19.17 & 112 &  83 & 0.57 \nl
 1239-12E &  12 45 32.21 &  12 51 58.68 & -12 43 21.7  & -11 16 50.5  &
 14.92 & 17.62 & 19.17 & 112 & 110 & 0.63 \nl
 1251-12W &  12 51 58.68 &  12 57 38.47 & -12 42 49.7  & -11 16 21.7  &
 15.03 & 17.73 & 19.38 & 112 &  93 & 0.47 \nl
 1251-12E &  12 57 38.47 &  13 04 18.02 & -12 42 49.7  & -11 16 21.7  &
 15.03 & 17.73 & 19.38 & 112 & 106 & 0.91 \nl
 1303-12W &  13 04 18.02 &  13 09 31.51 & -12 43 20.3  & -11 16 59.2  &
 14.97 & 17.67 & 19.22 & 112 &  68 & 0.87 \nl
 1303-12E &  13 09 31.51 &  13 15 59.11 & -12 43 20.3  & -11 16 59.2  &
 14.97 & 17.67 & 19.22 & 112 &  92 & 0.86 \nl
 1315-12W &  13 15 59.11 &  13 21 39.19 & -12 42 31.7  & -11 16 13.8  &
 14.92 & 17.62 & 19.27 & 112 &  57 & 0.76 \nl
 1315-12E &  13 21 39.19 &  13 28 17.59 & -12 42 31.7  & -11 16 13.8  &
 14.92 & 17.62 & 19.27 & 112 & 108 & 0.83 \nl
 1327-12W &  13 28 17.59 &  13 33 30.17 & -12 43 05.9  & -11 16 42.2  &
 15.05 & 17.75 & 19.30 & 112 &  96 & 0.50 \nl
 1327-12E &  13 33 30.17 &  13 39 58.15 & -12 43 05.9  & -11 16 42.2  &
 15.05 & 17.75 & 19.30 & 112 & 108 & 0.59 \nl
 1339-12W &  13 39 58.15 &  13 45 44.42 & -12 42 58.0  & -11 16 44.0  &
 14.87 & 17.57 & 19.22 & 112 &  89 & 0.81 \nl
 1339-12E &  13 45 44.42 &  13 52 17.78 & -12 42 58.0  & -11 16 44.0  &
 14.87 & 17.57 & 19.22 & 112 & 105 & 0.51 \nl
 1351-12W &  13 52 17.78 &  13 57 16.34 & -12 43 23.5  & -11 17 00.2  &
 14.95 & 17.65 & 19.20 & 112 &  93 & 0.57 \nl
 1351-12E &  13 57 16.34 &  14 04 01.10 & -12 43 23.5  & -11 17 00.2  &
 14.95 & 17.65 & 19.20 & 112 & 109 & 0.76 \nl
 1403-12W &  14 04 01.10 &  14 09 38.64 & -12 43 01.2  & -11 16 44.4  &
 14.98 & 17.68 & 19.33 & 112 &  94 & 0.66 \nl
 1403-12E &  14 09 38.64 &  14 16 16.68 & -12 43 01.2  & -11 16 44.4  &
 14.98 & 17.68 & 19.33 & 112 &  78 & 0.86 \nl
 1415-12W &  14 16 16.68 &  14 20 47.78 & -12 43 19.9  & -11 17 07.1  &
 14.97 & 17.77 & 19.27 & 112 &  42 & 0.93 \nl
 1415-12E &  14 20 47.78 &  14 27 00.82 & -12 43 19.9  & -11 17 07.1  &
 14.97 & 17.77 & 19.27 & 112 &  35 & 0.90 \nl
 1427-12W &  14 27 00.82 &  14 33 38.88 & -12 44 36.2  & -11 15 47.2  &
 14.89 & 17.69 & 18.79 & 112 &  66 & 0.81 \nl
 1427-12E &  14 33 38.88 &  14 40 15.60 & -12 44 36.2  & -11 15 47.2  &
 14.89 & 17.69 & 18.79 & 112 &  92 & 0.86 \nl
 1439-12W &  14 40 15.60 &  14 45 30.67 & -12 43 17.8  & -11 17 12.5  &
 15.02 & 17.92 & 19.37 & 112 &  70 & 0.63 \nl
 1439-12E &  14 45 30.67 &  14 51 57.58 & -12 43 17.8  & -11 17 12.5  &
 15.02 & 17.92 & 19.37 & 112 & 103 & 0.67 \nl
 1451-12W &  14 51 57.58 &  14 57 35.62 & -12 44 36.2  & -11 15 41.4  &
 14.80 & 17.60 & 18.70 & 112 &  62 & 0.86 \nl
 1451-12E &  14 57 35.62 &  15 04 16.87 & -12 44 36.2  & -11 15 41.4  &
 14.80 & 17.60 & 18.70 & 112 &  78 & 0.83 \nl
 1503-12W &  15 04 16.87 &  15 09 32.42 & -12 43 23.5  & -11 17 19.0  &
 15.04 & 17.94 & 19.39 & 112 &  90 & 0.69 \nl
 1503-12E &  15 09 32.42 &  15 15 57.79 & -12 43 23.5  & -11 17 19.0  &
 15.04 & 17.94 & 19.39 & 112 &  92 & 0.78 \nl
 1515-12W &  15 15 57.79 &  15 21 24.53 & -12 21 13.0  & -11 16 40.8  &
 14.92 & 17.92 & 19.42 & 112 &  44 & 0.72 \nl
 1515-12E &  15 21 24.53 &  15 27 50.71 & -12 21 13.0  & -11 16 40.8  &
 14.92 & 17.92 & 19.42 & 112 &  47 & 0.82 \nl
\tablebreak
 2100-39W &  21 00 40.22 &  21 07 58.73 & -39 46 18.8  & -38 16 13.8  &
 16.03 & 17.33 & 18.18 &  50 &  41 & 0.30 \nl
 2100-39E &  21 07 58.73 &  21 15 11.04 & -39 46 18.8  & -38 16 13.8  &
 15.03 & 17.54 & 18.48 & 112 &  79 & 0.54 \nl
 2114-39W &  21 15 11.04 &  21 23 55.27 & -39 44 43.1  & -38 16 16.7  &
 14.93 & 17.63 & 18.78 & 112 &  80 & 0.80 \nl
 2114-39M &  21 23 55.27 &  21 32 36.58 & -39 44 43.1  & -38 16 16.7  &
 14.93 & 17.63 & 18.78 & 112 &  89 & 0.68 \nl
 2114-39E &  21 32 36.58 &  21 41 11.35 & -39 44 43.1  & -38 16 16.7  &
 14.93 & 17.63 & 18.78 & 112 &  97 & 0.65 \nl
 2140-39W &  21 41 11.35 &  21 48 15.12 & -39 44 25.1  & -38 15 41.0  &
 15.95 & 17.25 & 18.10 &  50 &  32 & 0.75 \nl
 2140-39E &  21 48 15.12 &  21 55 09.46 & -39 44 25.1  & -38 15 41.0  &
 15.95 & 17.25 & 18.10 &  50 &  28 & 0.83 \nl
 2154-39W &  21 55 09.46 &  22 03 54.89 & -39 44 26.9  & -38 15 58.7  &
 14.95 & 17.65 & 18.80 & 112 &  67 & 0.74 \nl
 2154-39M &  22 03 54.89 &  22 12 40.30 & -39 44 26.9  & -38 15 58.7  &
 14.95 & 17.65 & 18.80 & 112 &  88 & 0.65 \nl
 2154-39E &  22 12 40.30 &  22 21 11.04 & -39 44 26.9  & -38 15 58.7  &
 14.95 & 17.65 & 18.80 & 112 & 108 & 0.59 \nl
 2220-39W &  22 21 11.04 &  22 28 12.58 & -39 44 25.1  & -38 15 25.9  &
 16.15 & 17.45 & 18.30 &  50 &  45 & 0.62 \nl
 2220-39E &  22 28 12.58 &  22 35 09.96 & -39 44 25.1  & -38 15 25.9  &
 16.15 & 17.45 & 18.30 &  50 &  44 & 0.35 \nl
 2234-39W &  22 35 09.96 &  22 43 52.30 & -39 44 55.7  & -38 16 17.0  &
 14.83 & 17.53 & 18.68 & 112 & 105 & 0.52 \nl
 2234-39M &  22 43 52.30 &  22 52 46.75 & -39 44 55.7  & -38 16 17.0  &
 14.83 & 17.53 & 18.68 & 112 & 103 & 0.83 \nl
 2234-39E &  22 52 46.75 &  23 01 11.83 & -39 44 55.7  & -38 16 17.0  &
 14.83 & 17.53 & 18.68 & 112 & 100 & 0.73 \nl
 2300-39W &  23 01 11.83 &  23 08 16.87 & -39 44 28.0  & -38 15 25.6  &
 16.03 & 17.33 & 18.18 &  50 &  37 & 0.64 \nl
 2300-39E &  23 08 16.87 &  23 15 11.21 & -39 44 28.0  & -38 15 25.6  &
 16.03 & 17.33 & 18.18 &  50 &  41 & 0.58 \nl
 2314-39W &  23 15 11.21 &  23 24 06.12 & -39 44 06.4  & -38 15 19.4  &
 14.98 & 17.68 & 18.83 & 112 &  91 & 0.83 \nl
 2314-39M &  23 24 06.12 &  23 32 38.35 & -39 44 06.4  & -38 15 19.4  &
 14.98 & 17.68 & 18.83 & 112 & 106 & 0.65 \nl
 2314-39E &  23 32 38.35 &  23 41 13.18 & -39 44 06.4  & -38 15 19.4  &
 14.98 & 17.68 & 18.83 & 112 &  99 & 0.58 \nl
 2340-39W &  23 41 13.18 &  23 48 15.58 & -39 44 23.6  & -38 15 07.6  &
 16.21 & 17.51 & 18.36 &  50 &  40 & 0.41 \nl
 2340-39E &  23 48 15.58 &  23 55 09.77 & -39 44 23.6  & -38 15 07.6  &
 16.21 & 17.51 & 18.36 &  50 &  45 & 0.41 \nl
 2354-39W &  23 55 09.77 &  00 04 01.22 & -39 44 26.5  & -38 15 25.9  &
 15.03 & 17.73 & 18.98 & 112 & 106 & 0.56 \nl
 2354-39M &  00 04 01.22 &  00 12 35.35 & -39 44 26.5  & -38 15 25.9  &
 15.03 & 17.73 & 18.98 & 112 & 105 & 0.69 \nl
 2354-39E &  00 12 35.35 &  00 21 12.12 & -39 44 26.5  & -38 15 25.9  &
 15.03 & 17.73 & 18.98 & 112 & 105 & 0.59 \nl
 0020-39W &  00 21 12.12 &  00 28 14.21 & -39 44 32.6  & -38 28 18.5  &
 16.12 & 17.42 & 18.27 &  50 &  36 & 0.46 \nl
 0020-39E &  00 28 14.21 &  00 35 10.54 & -39 44 32.6  & -38 28 18.5  &
 16.12 & 17.42 & 18.27 &  50 &  39 & 0.35 \nl
 0034-39W &  00 35 10.54 &  00 43 51.72 & -39 44 48.5  & -38 15 53.3  &
 15.03 & 17.73 & 18.88 & 112 & 107 & 0.74 \nl
 0034-39M &  00 43 51.72 &  00 52 42.79 & -39 44 48.5  & -38 15 53.3  &
 15.03 & 17.73 & 18.88 & 112 &  97 & 0.86 \nl
 0034-39E &  00 52 42.79 &  01 01 13.30 & -39 44 48.5  & -38 15 53.3  &
 15.03 & 17.73 & 18.88 & 112 &  93 & 0.83 \nl
 0100-39W &  01 01 13.30 &  01 08 21.22 & -39 44 19.3  & -38 15 19.8  &
 16.08 & 17.38 & 18.23 &  50 &  42 & 0.64 \nl
 0100-39E &  01 08 21.22 &  01 15 09.70 & -39 44 19.3  & -38 15 19.8  &
 16.08 & 17.38 & 18.23 &  50 &  39 & 0.45 \nl
 0114-39W &  01 15 09.70 &  01 23 51.65 & -39 43 48.4  & -38 15 16.2  &
 14.99 & 17.69 & 18.94 & 112 & 103 & 0.50 \nl
 0114-39M &  01 23 51.65 &  01 32 27.38 & -39 43 48.4  & -38 15 16.2  &
 14.99 & 17.69 & 18.94 & 112 &  97 & 0.64 \nl
 0114-39E &  01 32 27.38 &  01 41 11.40 & -39 43 48.4  & -38 15 16.2  &
 14.99 & 17.69 & 18.94 & 112 &  89 & 0.85 \nl
 0140-39W &  01 41 11.40 &  01 48 47.62 & -39 44 04.9  & -38 15 14.0  &
 16.14 & 17.44 & 18.29 &  50 &  24 & 1.00 \nl
 0140-39E &  01 48 47.62 &  01 55 09.72 & -39 44 04.9  & -38 15 14.0  &
 16.14 & 17.44 & 18.29 &  50 &  26 & 0.63 \nl
 0154-39W &  01 55 09.72 &  02 03 53.90 & -39 43 48.4  & -38 15 14.8  &
 14.92 & 17.62 & 18.77 & 112 &  88 & 0.62 \nl
 0154-39M &  02 03 53.90 &  02 12 43.75 & -39 43 48.4  & -38 15 14.8  &
 14.92 & 17.62 & 18.77 & 112 &  75 & 0.92 \nl
 0154-39E &  02 12 43.75 &  02 21 02.59 & -39 43 48.4  & -38 15 14.8  &
 14.92 & 17.62 & 18.77 & 112 &  86 & 0.89 \nl
 0220-39W &  02 21 02.59 &  02 28 13.27 & -39 44 09.6  & -38 15 05.8  &
 16.15 & 17.45 & 18.30 &  50 &  43 & 0.88 \nl
 0220-39E &  02 28 13.27 &  02 35 11.62 & -39 44 09.6  & -38 15 05.8  &
 16.15 & 17.45 & 18.30 &  50 &  47 & 0.83 \nl
 0234-39W &  02 35 11.62 &  02 43 52.44 & -39 43 38.6  & -38 15 13.3  &
 14.86 & 17.56 & 18.71 & 112 &  99 & 0.88 \nl
 0234-39M &  02 43 52.44 &  02 52 49.39 & -39 43 38.6  & -38 15 13.3  &
 14.86 & 17.56 & 18.71 & 112 &  78 & 0.81 \nl
 0234-39E &  02 52 49.39 &  03 01 10.68 & -39 43 38.6  & -38 15 13.3  &
 14.86 & 17.56 & 18.71 & 112 & 101 & 0.82 \nl
 0300-39W &  03 01 10.68 &  03 08 16.39 & -39 44 00.6  & -38 15 28.1  &
 16.02 & 17.32 & 18.27 &  50 &  36 & 0.31 \nl
 0300-39E &  03 08 16.39 &  03 15 09.34 & -39 44 00.6  & -38 15 28.1  &
 16.02 & 17.32 & 18.27 &  50 &  40 & 0.39 \nl
 0314-39W &  03 15 09.34 &  03 23 57.19 & -39 43 31.8  & -38 15 06.1  &
 15.02 & 17.72 & 18.87 & 112 & 105 & 0.61 \nl
 0314-39M &  03 23 57.19 &  03 32 42.67 & -39 43 31.8  & -38 15 06.1  &
 15.02 & 17.72 & 18.87 & 112 & 108 & 0.57 \nl
 0314-39E &  03 32 42.67 &  03 41 10.85 & -39 43 31.8  & -38 15 06.1  &
 15.02 & 17.72 & 18.87 & 112 & 107 & 0.44 \nl
 0340-39W &  03 41 10.85 &  03 48 17.93 & -39 44 10.7  & -38 15 16.9  &
 16.18 & 17.48 & 18.33 &  50 &  46 & 0.62 \nl
 0340-39E &  03 48 17.93 &  03 55 11.62 & -39 44 10.7  & -38 15 16.9  &
 16.18 & 17.48 & 18.33 &  50 &  40 & 0.46 \nl
 0354-39W &  03 55 11.62 &  04 03 53.64 & -39 43 18.1  & -38 15 20.5  &
 14.92 & 17.62 & 18.77 & 112 & 102 & 0.68 \nl
 0354-39M &  04 03 53.64 &  04 12 47.11 & -39 43 18.1  & -38 15 20.5  &
 14.92 & 17.62 & 18.77 & 112 &  82 & 0.82 \nl
 0354-39E &  04 12 47.11 &  04 21 11.88 & -39 43 18.1  & -38 15 20.5  &
 14.92 & 17.62 & 18.77 & 112 &  95 & 0.82 \nl
 0420-39W &  04 21 11.88 &  04 28 22.97 & -39 44 04.6  & -38 15 46.1  &
 16.07 & 17.37 & 18.22 &  50 &  31 & 0.49 \nl
 0420-39E &  04 28 22.97 &  04 35 49.25 & -39 44 04.6  & -38 15 46.1  &
 16.07 & 17.37 & 18.22 &  50 &  26 & 0.69 \nl
 2100-42W &  21 03 46.03 &  21 12 04.30 & -42 44 31.9  & -41 40 38.3  &
 14.79 & 17.59 & 18.69 & 112 & 103 & 0.80 \nl
 2100-42E &  21 12 04.30 &  21 20 22.70 & -42 44 31.9  & -41 40 38.3  &
 14.79 & 17.59 & 18.69 & 112 &  80 & 0.85 \nl
 2120-42W &  21 20 43.92 &  21 28 49.03 & -42 44 55.3  & -41 16 05.5  &
 14.89 & 17.69 & 18.79 & 112 &  84 & 0.63 \nl
 2120-42E &  21 28 49.03 &  21 35 44.57 & -42 44 55.3  & -41 16 05.5  &
 14.89 & 17.69 & 18.79 & 112 &  80 & 0.52 \nl
 2135-42W &  21 35 44.57 &  21 43 45.19 & -42 44 19.3  & -41 15 30.2  &
 14.82 & 17.52 & 18.82 & 112 & 102 & 0.84 \nl
 2135-42M &  21 43 45.19 &  21 52 50.28 & -42 44 19.3  & -41 15 30.2  &
 14.82 & 17.52 & 18.82 & 112 &  84 & 0.85 \nl
 2135-42E &  21 52 50.28 &  22 01 22.10 & -42 44 19.3  & -41 15 30.2  &
 14.82 & 17.52 & 18.82 & 112 &  77 & 0.81 \nl
 2200-42W &  22 01 22.10 &  22 08 40.54 & -42 45 11.2  & -41 16 20.6  &
 14.76 & 17.56 & 18.66 & 112 &  73 & 0.91 \nl
 2200-42E &  22 08 40.54 &  22 15 43.92 & -42 45 11.2  & -41 16 20.6  &
 14.76 & 17.56 & 18.66 & 112 &  53 & 0.78 \nl
 2215-42W &  22 15 43.92 &  22 23 51.84 & -42 44 02.4  & -41 14 47.0  &
 14.83 & 17.53 & 18.83 & 112 & 102 & 0.83 \nl
 2215-42M &  22 23 51.84 &  22 32 57.19 & -42 44 02.4  & -41 14 47.0  &
 14.83 & 17.53 & 18.83 & 112 & 106 & 0.67 \nl
 2215-42E &  22 32 57.19 &  22 41 22.82 & -42 44 02.4  & -41 14 47.0  &
 14.83 & 17.53 & 18.83 & 112 & 106 & 0.65 \nl
 2240-42W &  22 41 22.82 &  22 48 47.59 & -42 45 13.0  & -41 16 05.2  &
 14.92 & 17.72 & 19.02 & 112 &  89 & 0.68 \nl
 2240-42E &  22 48 47.59 &  22 56 43.73 & -42 45 13.0  & -41 16 05.2  &
 14.92 & 17.72 & 19.02 & 112 & 101 & 0.92 \nl
 2255-42W &  22 56 43.73 &  23 03 40.49 & -42 44 43.4  & -41 15 18.4  &
 14.84 & 17.54 & 18.59 & 112 &  78 & 0.77 \nl
 2255-42M &  23 03 40.49 &  23 12 57.48 & -42 44 43.4  & -41 15 18.4  &
 14.84 & 17.54 & 18.59 & 112 & 101 & 0.82 \nl
 2255-42E &  23 12 57.48 &  23 21 19.94 & -42 44 43.4  & -41 15 18.4  &
 14.84 & 17.54 & 18.59 & 112 & 106 & 0.34 \nl
 2320-42W &  23 21 19.94 &  23 28 33.14 & -42 44 38.4  & -41 15 31.7  &
 16.26 & 17.56 & 18.41 &  50 &  37 & 0.58 \nl
 2320-42E &  23 28 33.14 &  23 35 44.54 & -42 44 38.4  & -41 15 31.7  &
 16.26 & 17.56 & 18.41 &  50 &  39 & 0.95 \nl
 2335-42W &  23 35 44.54 &  23 43 43.99 & -42 44 20.8  & -41 14 53.2  &
 15.04 & 17.74 & 18.99 & 112 & 101 & 0.76 \nl
 2335-42M &  23 43 43.99 &  23 52 56.16 & -42 44 20.8  & -41 14 53.2  &
 15.04 & 17.74 & 18.99 & 112 &  99 & 0.90 \nl
 2335-42E &  23 52 56.16 &  00 01 25.73 & -42 44 20.8  & -41 14 53.2  &
 15.04 & 17.74 & 18.99 & 112 & 103 & 0.65 \nl
 0000-42W &  00 01 25.73 &  00 08 28.66 & -42 44 04.6  & -41 15 28.8  &
 16.19 & 17.49 & 18.34 &  50 &  40 & 0.35 \nl
 0000-42E &  00 08 28.66 &  00 15 43.22 & -42 44 04.6  & -41 15 28.8  &
 16.19 & 17.49 & 18.34 &  50 &  37 & 0.31 \nl
 0015-42W &  00 15 43.22 &  00 23 41.14 & -42 44 13.9  & -41 14 42.7  &
 14.93 & 17.63 & 18.78 & 112 & 105 & 0.56 \nl
 0015-42M &  00 23 41.14 &  00 33 03.82 & -42 44 13.9  & -41 14 42.7  &
 14.93 & 17.63 & 18.78 & 112 & 107 & 0.85 \nl
 0015-42E &  00 33 03.82 &  00 41 21.14 & -42 44 13.9  & -41 14 42.7  &
 14.93 & 17.63 & 18.78 & 112 &  81 & 0.88 \nl
 0040-42W &  00 41 21.14 &  00 48 29.50 & -42 44 00.2  & -41 15 06.8  &
 16.22 & 17.52 & 18.37 &  50 &  47 & 0.64 \nl
 0040-42E &  00 48 29.50 &  00 55 46.30 & -42 44 00.2  & -41 15 06.8  &
 16.22 & 17.52 & 18.37 &  50 &  49 & 0.72 \nl
 0054-42W &  00 55 46.30 &  01 04 26.52 & -42 44 03.1  & -41 27 16.2  &
 15.01 & 17.81 & 18.91 & 112 &  99 & 0.93 \nl
 0054-42M &  01 04 26.52 &  01 12 58.32 & -42 44 03.1  & -41 27 16.2  &
 15.01 & 17.81 & 18.91 & 112 & 103 & 0.87 \nl
 0054-42E &  01 12 58.32 &  01 21 29.71 & -42 44 03.1  & -41 27 16.2  &
 15.01 & 17.81 & 18.91 & 112 &  91 & 0.90 \nl
 0120-42W &  01 21 29.71 &  01 28 38.81 & -42 44 04.6  & -41 15 17.3  &
 16.16 & 17.46 & 18.31 &  50 &  47 & 0.63 \nl
 0120-42E &  01 28 38.81 &  01 35 46.56 & -42 44 04.6  & -41 15 17.3  &
 16.16 & 17.46 & 18.31 &  50 &  47 & 0.55 \nl
 0134-42W &  01 35 46.56 &  01 44 20.52 & -42 43 50.5  & -41 14 52.1  &
 14.99 & 17.79 & 18.89 & 112 & 107 & 0.62 \nl
 0134-42M &  01 44 20.52 &  01 52 54.14 & -42 43 50.5  & -41 14 52.1  &
 14.99 & 17.79 & 18.89 & 112 &  87 & 0.91 \nl
 0134-42E &  01 52 54.14 &  02 01 22.44 & -42 31 04.8  & -41 14 52.1  &
 14.99 & 17.79 & 18.89 & 112 &  69 & 0.93 \nl
 0200-42W &  02 01 40.13 &  02 08 48.19 & -42 43 53.0  & -41 15 40.7  &
 16.33 & 17.63 & 18.58 &  50 &  40 & 0.39 \nl
 0200-42E &  02 08 48.19 &  02 16 29.81 & -42 43 53.0  & -41 15 40.7  &
 16.33 & 17.63 & 18.58 &  50 &  41 & 0.55 \nl
 0214-42W &  02 18 20.95 &  02 24 20.42 & -42 42 55.1  & -41 14 12.5  &
 15.00 & 17.80 & 18.90 & 112 &  58 & 0.71 \nl
 0214-42M &  02 24 20.42 &  02 32 58.78 & -42 42 55.1  & -41 14 12.5  &
 15.00 & 17.80 & 18.90 & 112 &  90 & 0.84 \nl
 0214-42E &  02 32 58.78 &  02 41 20.26 & -42 42 55.1  & -41 14 12.5  &
 15.00 & 17.80 & 18.90 & 112 &  73 & 0.56 \nl
 0240-42W &  02 41 20.26 &  02 48 35.23 & -42 44 02.4  & -41 15 03.2  &
 16.23 & 17.53 & 18.48 &  50 &  39 & 0.28 \nl
 0240-42E &  02 48 35.23 &  02 55 46.03 & -42 44 02.4  & -41 15 03.2  &
 16.23 & 17.53 & 18.48 &  50 &  42 & 0.48 \nl
 0254-42W &  02 55 46.03 &  03 04 05.11 & -42 43 23.9  & -41 14 48.8  &
 15.09 & 17.89 & 18.99 & 112 & 110 & 0.79 \nl
 0254-42M &  03 04 05.11 &  03 13 08.71 & -42 43 23.9  & -41 14 48.8  &
 15.09 & 17.89 & 18.99 & 112 & 102 & 0.73 \nl
 0254-42E &  03 13 08.71 &  03 21 22.25 & -42 43 23.9  & -41 14 48.8  &
 15.09 & 17.89 & 18.99 & 112 & 106 & 0.37 \nl
 0320-42W &  03 21 22.25 &  03 28 41.09 & -42 44 14.3  & -41 15 04.3  &
 16.26 & 17.56 & 18.41 &  50 &  34 & 0.51 \nl
 0320-42E &  03 28 41.09 &  03 35 44.62 & -42 44 14.3  & -41 15 04.3  &
 16.26 & 17.56 & 18.41 &  50 &  43 & 0.51 \nl
 0335-42W &  03 35 44.62 &  03 43 40.32 & -42 43 14.9  & -41 27 20.9  &
 14.95 & 17.65 & 18.80 & 112 &  76 & 0.88 \nl
 0335-42M &  03 43 40.32 &  03 52 59.26 & -42 43 14.9  & -41 27 20.9  &
 14.95 & 17.65 & 18.80 & 112 &  93 & 0.91 \nl
 0335-42E &  03 52 59.26 &  04 01 28.42 & -42 43 14.9  & -41 27 20.9  &
 14.95 & 17.65 & 18.80 & 112 &  96 & 0.88 \nl
 0400-42W &  04 01 28.42 &  04 08 47.14 & -42 44 03.1  & -41 15 13.0  &
 15.00 & 17.70 & 18.85 & 112 &  73 & 0.77 \nl
 0400-42E &  04 08 47.14 &  04 15 47.04 & -42 44 03.1  & -41 15 13.0  &
 15.00 & 17.80 & 18.90 & 112 &  74 & 0.78 \nl
 0415-42W &  04 15 47.04 &  04 24 48.46 & -42 43 04.1  & -41 14 37.0  &
 14.87 & 17.57 & 18.72 & 112 & 102 & 0.83 \nl
 0415-42E &  04 24 48.46 &  04 33 44.47 & -42 43 04.1  & -41 14 37.0  &
 14.87 & 17.57 & 18.72 & 112 &  61 & 0.79 \nl
 2100-45W &  21 00 45.60 &  21 09 09.60 & -45 45 00.0  & -44 16 12.0  &
 14.93 & 17.63 & 18.88 & 112 &  96 & 0.64 \nl
 2100-45E &  21 09 09.60 &  21 17 33.60 & -45 45 00.0  & -44 16 12.0  &
 14.93 & 17.63 & 18.88 & 112 &  99 & 0.41 \nl
 2115-45W &  21 17 33.60 &  21 25 02.40 & -45 43 48.0  & -44 15 36.0  &
 15.00 & 17.70 & 19.00 & 112 &  75 & 0.78 \nl
 2115-45E &  21 33 38.40 &  21 42 40.80 & -45 43 48.0  & -44 15 36.0  &
 15.00 & 17.70 & 19.00 & 112 & 101 & 0.66 \nl
 2140-45W &  21 42 40.80 &  21 47 55.20 & -45 46 48.0  & -44 16 12.0  &
 16.16 & 17.46 & 18.31 &  50 &  27 & 0.37 \nl
 2140-45E &  21 47 55.20 &  21 55 43.20 & -45 46 48.0  & -44 16 12.0  &
 16.16 & 17.46 & 18.31 &  50 &  38 & 0.53 \nl
 2155-45W &  21 55 43.20 &  22 05 02.40 & -45 43 48.0  & -44 15 36.0  &
 14.94 & 17.64 & 18.94 & 112 &  76 & 0.60 \nl
 2155-45M &  22 05 02.40 &  22 13 48.00 & -45 43 48.0  & -44 15 36.0  &
 14.94 & 17.64 & 18.94 & 112 &  83 & 0.62 \nl
 2155-45E &  22 13 48.00 &  22 20 50.40 & -45 43 48.0  & -44 15 36.0  &
 14.94 & 17.64 & 18.94 & 112 &  79 & 0.83 \nl
 2220-45W &  22 20 50.40 &  22 29 12.00 & -45 44 24.0  & -44 15 36.0  &
 14.98 & 17.78 & 19.08 & 112 &  90 & 0.83 \nl
 2220-45E &  22 29 12.00 &  22 37 36.00 & -45 44 24.0  & -44 15 36.0  &
 14.98 & 17.78 & 19.08 & 112 &  90 & 0.61 \nl
 2235-45W &  22 37 36.00 &  22 45 09.60 & -45 44 24.0  & -44 15 00.0  &
 15.06 & 17.76 & 19.06 & 112 &  85 & 0.44 \nl
 2235-45M &  22 45 09.60 &  22 53 52.80 & -45 44 24.0  & -44 15 00.0  &
 15.06 & 17.76 & 19.06 & 112 & 108 & 0.52 \nl
 2235-45E &  22 53 52.80 &  23 02 38.40 & -45 44 24.0  & -44 15 00.0  &
 15.06 & 17.76 & 19.06 & 112 & 105 & 0.47 \nl
 2300-45W &  23 02 38.40 &  23 09 00.00 & -45 44 24.0  & -44 15 00.0  &
 16.18 & 17.48 & 18.33 &  50 &  25 & 0.29 \nl
 2300-45E &  23 09 00.00 &  23 15 45.60 & -45 44 24.0  & -44 40 12.0  &
 16.18 & 17.48 & 18.33 &  50 &  30 & 0.52 \nl
 2315-45W &  23 15 45.60 &  23 25 02.40 & -45 44 24.0  & -44 15 36.0  &
 15.02 & 17.72 & 19.02 & 112 &  89 & 0.71 \nl
 2315-45M &  23 25 02.40 &  23 33 36.00 & -45 44 24.0  & -44 15 36.0  &
 15.02 & 17.72 & 19.02 & 112 &  74 & 0.68 \nl
 2315-45E &  23 33 36.00 &  23 40 43.20 & -45 44 24.0  & -44 15 36.0  &
 15.02 & 17.72 & 19.02 & 112 &  87 & 0.87 \nl
 2340-45W &  23 40 43.20 &  23 48 57.60 & -45 44 24.0  & -44 15 36.0  &
 16.18 & 17.48 & 18.43 &  50 &  40 & 0.68 \nl
 2340-45E &  23 48 57.60 &  23 55 45.60 & -45 44 24.0  & -44 15 36.0  &
 15.18 & 17.69 & 18.63 & 112 &  68 & 0.81 \nl
 2355-45W &  23 55 45.60 &  00 05 09.60 & -45 44 24.0  & -44 15 00.0  &
 14.91 & 17.61 & 18.91 & 112 & 103 & 0.74 \nl
 2355-45M &  00 05 09.60 &  00 13 36.00 & -45 44 24.0  & -44 15 00.0  &
 14.91 & 17.61 & 18.91 & 112 &  95 & 0.91 \nl
 2355-45E &  00 13 36.00 &  00 22 36.00 & -45 44 24.0  & -44 15 00.0  &
 14.91 & 17.61 & 18.91 & 112 &  72 & 0.80 \nl
 0020-45W &  00 22 36.00 &  00 29 09.60 & -45 44 24.0  & -44 15 36.0  &
 16.21 & 17.51 & 18.36 &  50 &  36 & 0.92 \nl
 0020-45E &  00 29 09.60 &  00 37 12.00 & -45 44 24.0  & -44 15 36.0  &
 16.21 & 17.51 & 18.36 &  50 &  38 & 0.83 \nl
 0035-45W &  00 37 12.00 &  00 45 04.80 & -45 44 24.0  & -44 15 00.0  &
 14.84 & 17.54 & 18.84 & 112 &  81 & 0.66 \nl
 0035-45M &  00 45 04.80 &  00 53 26.40 & -45 44 24.0  & -44 15 00.0  &
 14.84 & 17.54 & 18.84 & 112 &  74 & 0.71 \nl
 0035-45E &  00 53 26.40 &  01 00 52.80 & -45 44 24.0  & -44 15 00.0  &
 14.84 & 17.54 & 18.84 & 112 &  68 & 0.85 \nl
 0100-45W &  01 00 52.80 &  01 09 16.80 & -45 45 00.0  & -44 16 12.0  &
 15.09 & 17.79 & 19.14 & 112 &  87 & 0.89 \nl
 0100-45E &  01 09 16.80 &  01 17 45.60 & -45 45 00.0  & -44 16 12.0  &
 15.09 & 17.79 & 19.14 & 112 & 109 & 0.65 \nl
 0115-45W &  01 17 45.60 &  01 24 09.60 & -45 43 48.0  & -44 15 00.0  &
 14.94 & 17.64 & 18.94 & 112 &  82 & 0.54 \nl
 0115-45M &  01 24 09.60 &  01 33 38.40 & -45 43 48.0  & -44 15 00.0  &
 14.94 & 17.64 & 18.94 & 112 & 103 & 0.59 \nl
 0115-45E &  01 33 38.40 &  01 42 04.80 & -45 43 48.0  & -44 15 00.0  &
 14.94 & 17.64 & 18.94 & 112 &  99 & 0.79 \nl
 0140-45W &  01 42 04.80 &  01 49 02.40 & -45 44 24.0  & -44 15 36.0  &
 16.17 & 17.47 & 18.32 &  50 &  35 & 0.57 \nl
 0140-45E &  01 49 02.40 &  01 55 48.00 & -45 44 24.0  & -44 15 36.0  &
 16.17 & 17.47 & 18.32 &  50 &  35 & 0.66 \nl
 0155-45W &  01 55 48.00 &  02 04 07.20 & -45 43 48.0  & -44 15 00.0  &
 14.94 & 17.64 & 18.94 & 112 & 103 & 0.75 \nl
 0155-45M &  02 04 07.20 &  02 13 55.20 & -45 43 48.0  & -44 15 00.0  &
 14.94 & 17.64 & 18.94 & 112 & 104 & 0.68 \nl
 0155-45E &  02 13 55.20 &  02 22 40.80 & -45 43 48.0  & -44 15 00.0  &
 14.94 & 17.64 & 18.94 & 112 & 105 & 0.78 \nl
 0220-45W &  02 22 40.80 &  02 29 07.20 & -45 43 48.0  & -44 15 00.0  &
 16.06 & 17.36 & 18.21 &  50 &  35 & 0.61 \nl
 0220-45E &  02 29 07.20 &  02 35 48.00 & -45 43 48.0  & -44 15 00.0  &
 16.06 & 17.36 & 18.21 &  50 &  32 & 0.81 \nl
 0235-45W &  02 35 48.00 &  02 44 04.80 & -45 43 12.0  & -44 14 24.0  &
 14.89 & 17.59 & 18.89 & 112 & 107 & 0.82 \nl
 0235-45M &  02 44 04.80 &  02 53 45.60 & -45 43 12.0  & -44 14 24.0  &
 14.89 & 17.59 & 18.89 & 112 & 107 & 0.78 \nl
 0235-45E &  02 53 45.60 &  03 02 40.80 & -45 43 12.0  & -44 14 24.0  &
 14.89 & 17.59 & 18.89 & 112 &  83 & 0.84 \nl
 0300-45W &  03 02 40.80 &  03 09 19.20 & -45 43 48.0  & -44 15 36.0  &
 16.08 & 17.38 & 18.23 &  50 &  41 & 0.66 \nl
 0300-45E &  03 09 19.20 &  03 17 19.20 & -45 43 48.0  & -44 15 36.0  &
 16.08 & 17.38 & 18.23 &  50 &  48 & 0.21 \nl
 0315-45E &  03 33 40.80 &  03 42 38.40 & -45 43 12.0  & -44 15 00.0  &
 14.89 & 17.59 & 18.89 & 112 & 101 & 0.77 \nl
 0340-45W &  03 42 38.40 &  03 49 02.40 & -45 43 48.0  & -44 15 36.0  &
 16.05 & 17.35 & 18.30 &  50 &  36 & 0.49 \nl
 0340-45E &  03 49 02.40 &  03 55 45.60 & -45 43 48.0  & -44 15 36.0  &
 16.05 & 17.35 & 18.30 &  50 &  34 & 0.43 \nl
 0355-45W &  03 55 45.60 &  04 05 04.80 & -45 43 12.0  & -44 15 36.0  &
 14.82 & 17.52 & 18.82 & 112 & 105 & 0.67 \nl
 0355-45M &  04 05 04.80 &  04 13 24.00 & -45 43 12.0  & -44 15 36.0  &
 14.82 & 17.52 & 18.82 & 112 &  60 & 0.83 \nl
 0355-45E &  04 13 24.00 &  04 22 38.40 & -45 43 12.0  & -44 15 36.0  &
 14.82 & 17.52 & 18.82 & 112 & 103 & 0.78 \nl
 0420-45W &  04 22 38.40 &  04 29 02.40 & -45 43 48.0  & -44 15 36.0  &
 16.14 & 17.44 & 18.29 &  50 &  32 & 0.85 \nl
 0420-45E &  04 29 02.40 &  04 37 19.20 & -45 43 48.0  & -44 15 36.0  &
 16.14 & 17.44 & 18.29 &  50 &  43 & 0.66 \nl
\enddata
\tablenotetext{a}{$\alpha_1$ and $\alpha_2$ denote the 
                  right ascension limits of each field,
                  and $\delta_1$ and $\delta_2$ denote the 
                  declination limits. All coordinates are epoch 1950.0.}
\tablenotetext{b}{$m_1$ and $m_2$ denote the isophotal magnitude limits of
                  each field, and $m_{cen}$ is the faint central magnitude
                  limit at $m_2$. The isophotal magnitude $m$ and the 
                  central magnitude $m_c$ of each galaxy must meet the 
                  photometric selection criteria
                  $m_1 \leq m < m_2$ and $m_c < m_{cen} - 0.5 (m_2 - m)$.}
\tablenotetext{c}{$N_{fib}$ denotes whether data for that field was
                  obtained with the 50-fiber or the 112-fiber spectrograph
                  system.}
\tablenotetext{d}{$N_{gal}$ denotes the number of galaxy redshifts observed 
                  in each field.}
\tablenotetext{e}{$f$ denotes the galaxy sampling fraction for each field.}
\end{deluxetable}

\clearpage

\clearpage

\clearpage

\clearpage

\clearpage

\clearpage

\clearpage

\clearpage

\begin{figure}
\caption{(a) The distribution of galaxies for the slice centered at
       $\delta = -3 \arcdeg$. The galaxy distribution is shown as a
       function of heliocentric velocity and right ascension in the 
       main portion of the plot, and as a function of declination and 
       right ascension (epoch 1950.0) in
       the upper circular part of the figure.}
\label{fig-3}
\end{figure}


\begin{figure}
\figurenum{8}
\caption{(b) The distribution of galaxies for the slice centered at
       $\delta = -6 \arcdeg$.}
\label{fig-6}
\end{figure}


\begin{figure}
\figurenum{8}
\caption{(c) The distribution of galaxies for the slice centered at
       $\delta = -12 \arcdeg$.}
\label{fig-12}
\end{figure}


\begin{figure}
\figurenum{8}
\caption{(d) The distribution of galaxies for the slice centered at
       $\delta = -39 \arcdeg$.}
\label{fig-39}
\end{figure}


\begin{figure}
\figurenum{8}
\caption{(e) The distribution of galaxies for the slice centered at
       $\delta = -42 \arcdeg$.}
\label{fig-42}
\end{figure}


\begin{figure}
\figurenum{8}
\caption{(f) The distribution of galaxies for the slice centered at
       $\delta = -45 \arcdeg$.}
\label{fig-45}
\end{figure}

\clearpage

\begin{figure}
\figurenum{8}
\plotfiddle{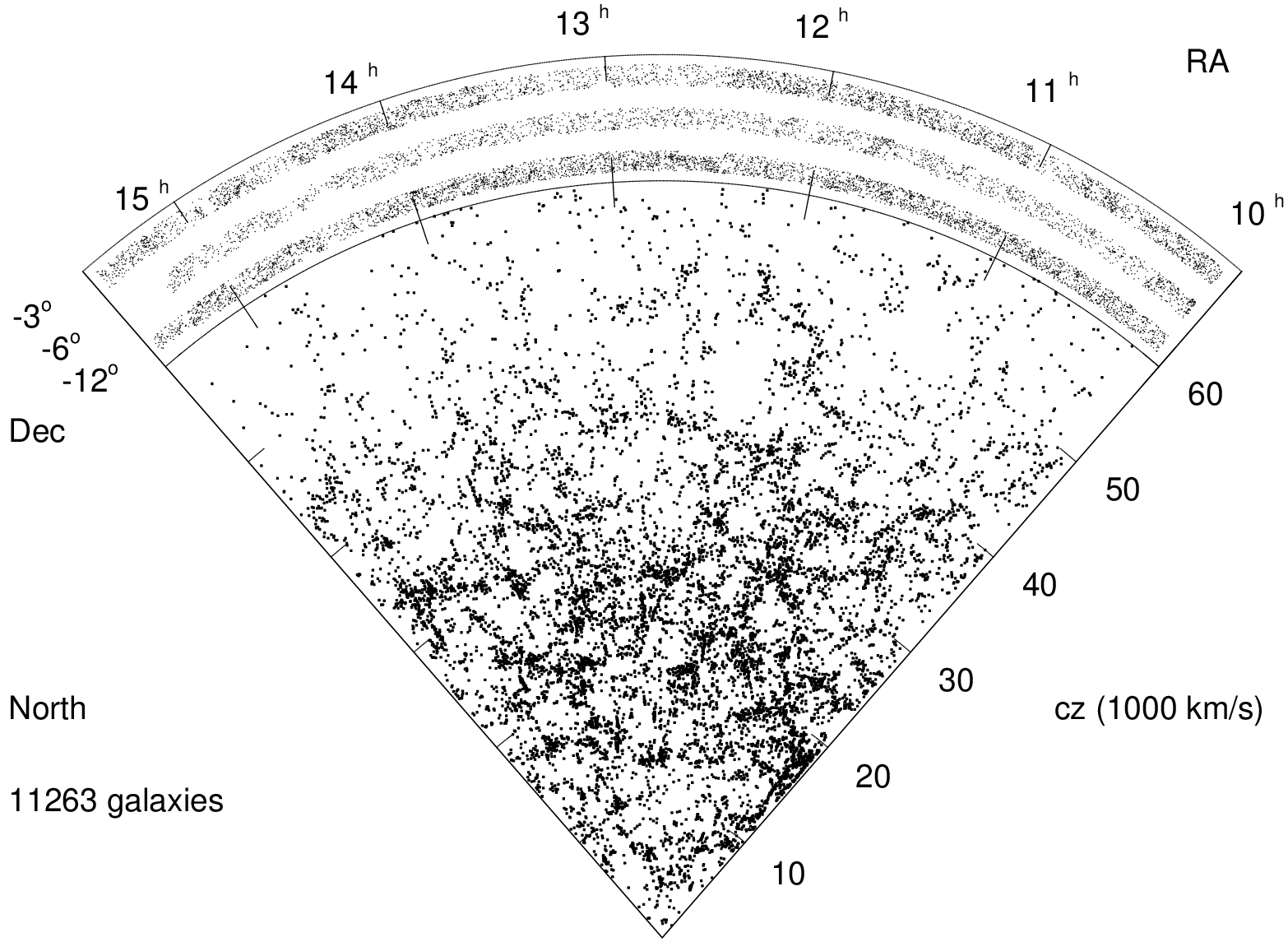}{13cm}{0}{62}{62}{-250}{0}
\caption{(g) The distribution of galaxies for all three slices
       in the North galactic cap.}
\label{fignor}
\end{figure}

\clearpage

\begin{figure}
\figurenum{8}
\plotfiddle{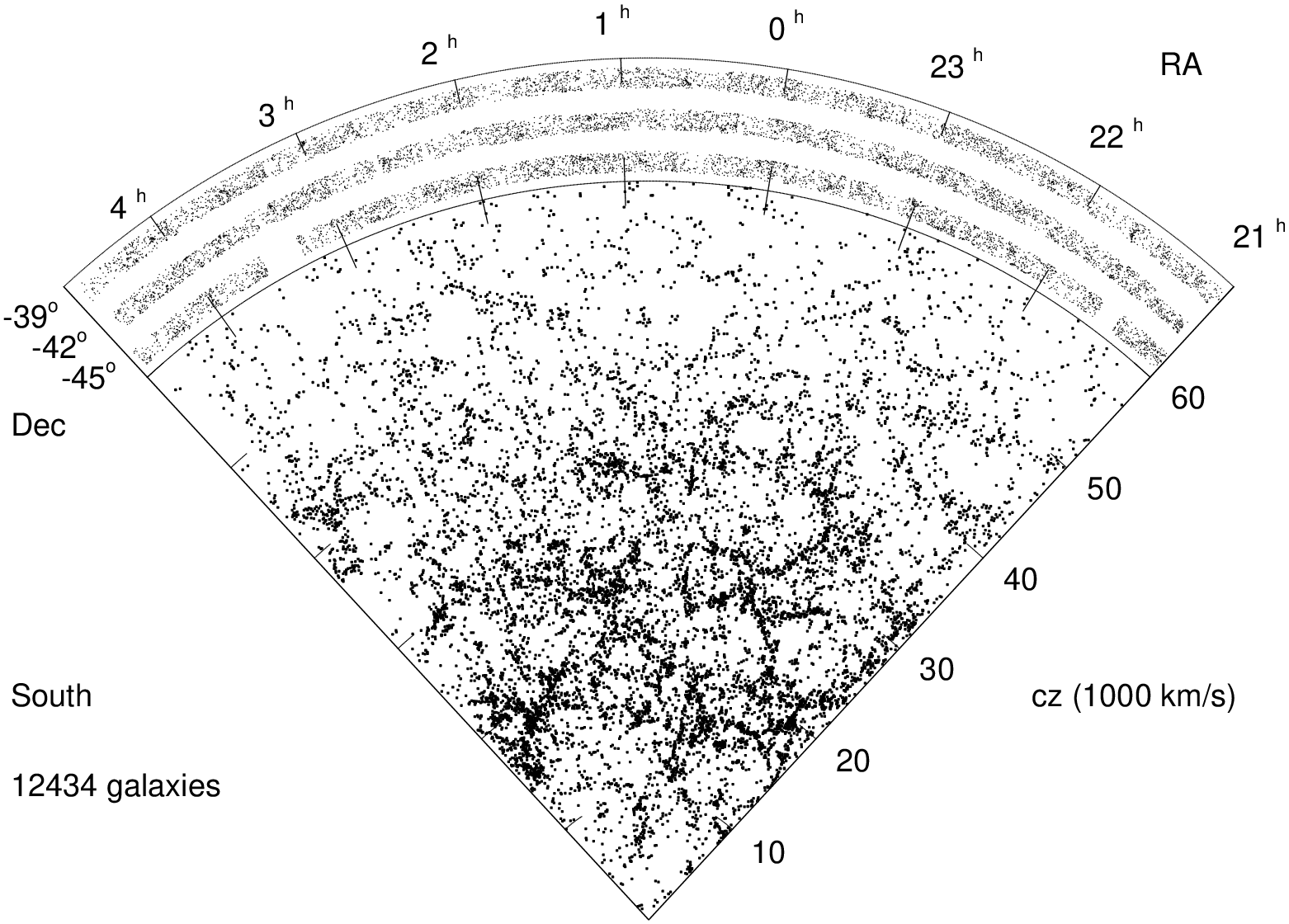}{13cm}{0}{62}{62}{-250}{0}
\caption{(h) The distribution of galaxies for all three slices
       in the South galactic cap.}
\label{figsou}
\end{figure}

\clearpage

\begin{figure}
\plotone{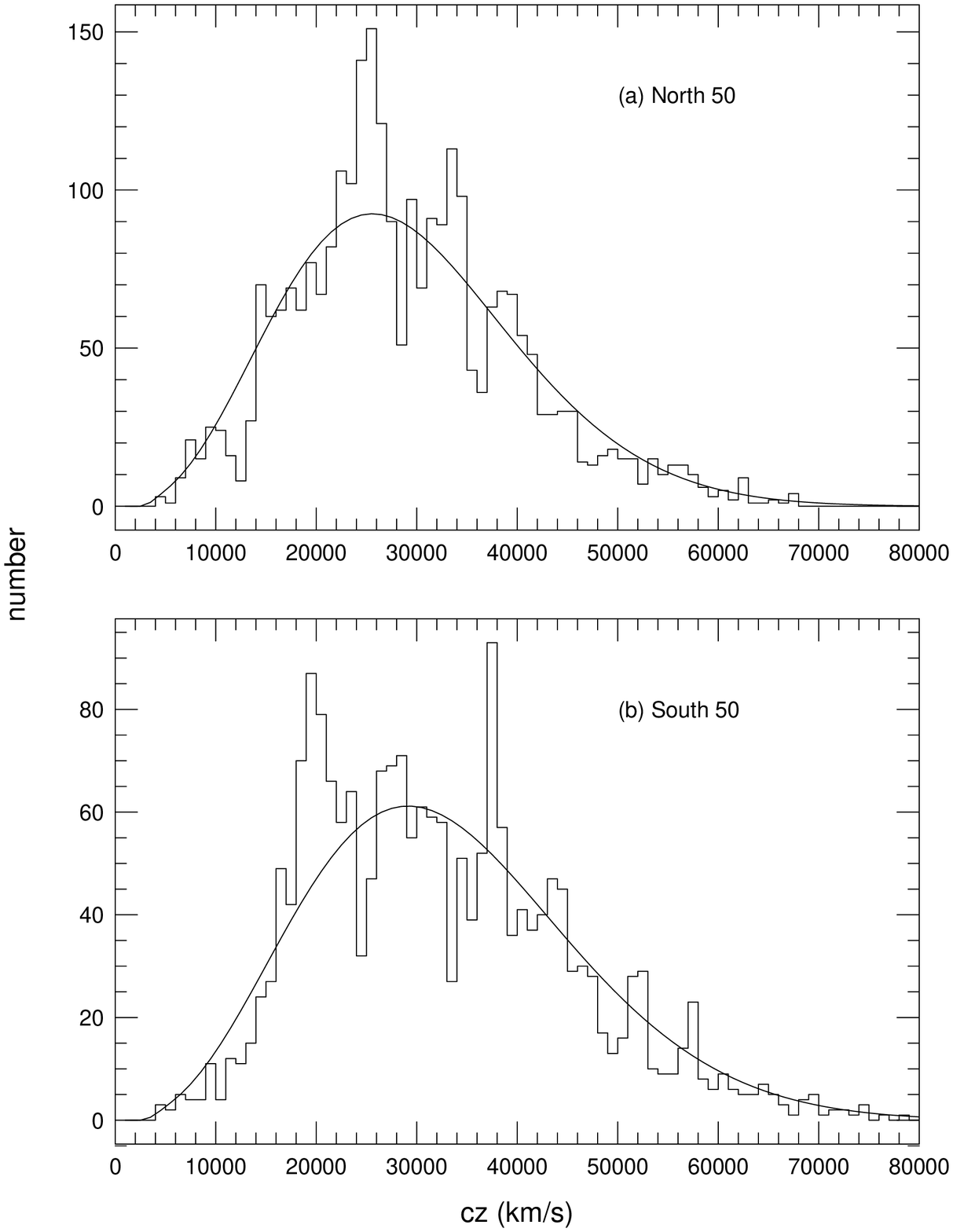}
\caption{The redshift histogram for the (a) North and (b) South 50-fiber
   samples. The smooth lines show the expected distribution 
   if galaxies were uniformly distributed with the LCRS luminosity
   function.}
\label{figh50}
\end{figure}

\clearpage

\begin{figure}
\plotone{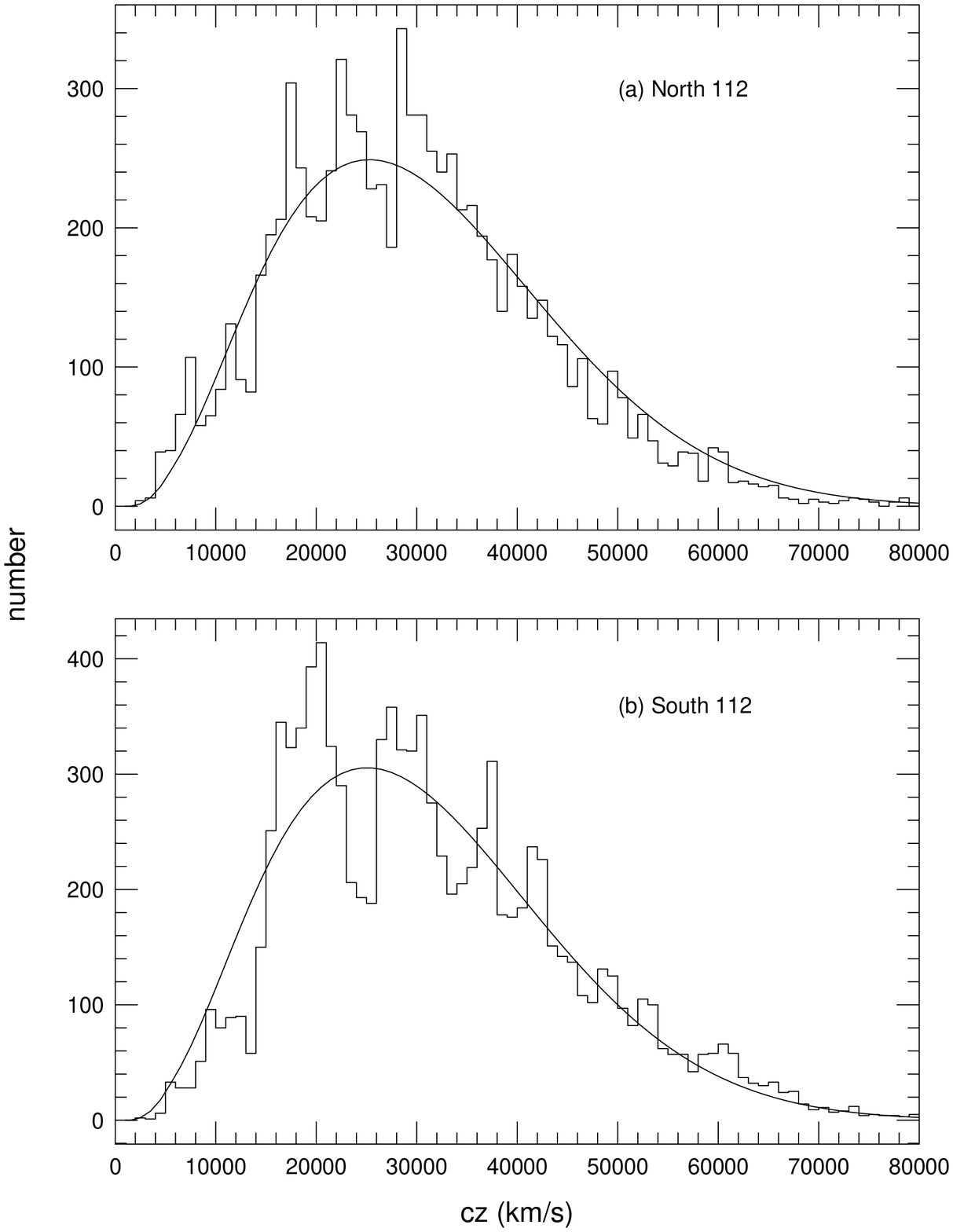}
\caption{The redshift histogram for the (a) North and (b) South 112-fiber
   samples. The smooth lines show the expected distribution 
   if galaxies were uniformly distributed with the LCRS luminosity
   function.}
\label{figh112}
\end{figure}

\end{document}